\documentclass[journal]{vgtc}                     

\onlineid{0}

\vgtccategory{Research}

\vgtcpapertype{Data Transformations}

\title{Visual analysis of bivariate dependence between continuous random variables}

\author{%
  \authororcid{Arturo Erdely}{0000-0003-1653-8342} and \authororcid{Manuel Rubio-S\'{a}nchez}{0000-0002-8692-2553}
}

\authorfooter{
  \item
  	Arturo Erdely. Universidad Nacional Aut\'{o}noma de M\'{e}xico.
    E-mail: aerdely@acatlan.unam.mx.
  \item
  	Manuel Rubio-S\'{a}nchez. Universidad Rey Juan Carlos.
    E-mail: manuel.rubio@urjc.es.
}

\abstract{%
Scatter plots are widely recognized as fundamental tools for illustrating the relationship between two numerical variables. Despite this, based on solid theoretical foundations, scatter plots generated from pairs of continuous random variables may not serve as reliable tools for assessing dependence. Sklar's Theorem implies that scatter plots created from ranked data are preferable for such analysis as they exclusively convey information pertinent to dependence. This is in stark contrast to conventional scatter plots, which also encapsulate information about the variables' marginal distributions. Such additional information is extraneous to dependence analysis and can obscure the visual interpretation of the variables' relationship. In this article, we delve into the theoretical underpinnings of these ranked data scatter plots, hereafter referred to as rank plots. We offer insights into interpreting the information they reveal and examine their connections with various association measures, including Pearson’s and Spearman’s correlation coefficients, as well as Schweizer-Wolff's measure of dependence. Furthermore, we introduce a novel graphical combination for dependence analysis, termed a \textit{dplot,} and demonstrate its efficacy through real data examples.}


\usepackage{tabu}                      
\usepackage{booktabs}                  

\usepackage{mathptmx}                  
\usepackage{amsmath}
\usepackage{amsfonts,amsbsy}
\usepackage{amssymb}
\usepackage{rotating}
\usepackage{psfrag}
\usepackage{array}
\usepackage{url}
\usepackage{float}

\usepackage[usenames]{color}
\definecolor{darkblue}{rgb}{0,0,0.75}
\definecolor{darkgreen}{rgb}{0,0.5,0}
\definecolor{darkred}{rgb}{0.8,0,0}
\definecolor{darkgray}{rgb}{0.3,0.3,0.3}

\begin{document}


\firstsection{Introduction}

\maketitle


\label{sec:introduction}


For over three decades, the data visualization community has been innovating, crafting sophisticated and interactive methods to analyze and present data. Despite these advancements, simple visual representations such as bar, line, and pie charts remain indispensable. Their simplicity not only facilitates communication with a wide audience, they can also be considered to be the most appropriate and effective visualizations of certain types of information.

Among these foundational visual tools is the scatter plot, which is regarded as one of the most useful and popular statistical graphs \cite{friendly05}. It depicts bivariate numerical data as points within a Cartesian coordinate system, presenting the data ``as it is'' (i.e., no information is lost through its visual encoding), allowing for the direct reading of values by projecting the points onto labeled axes. In this regard, scatter plots have numerous benefits and are the predominant technique for visualizing data of two numerical variables simultaneously. They can be extended by incorporating additional layers, such as regression curves, confidence bands, modifications in point characteristics (shape, size, color, opacity), regions/areas of interest, histograms, density contours, correlation/dependence measures, winglets, glyphs, and so forth (see \cite{Friendly06,Elmqvist08b,Chan13,Yates14,Nguyen16,Nguyen17,Sarikaya18,Lu20}). Moreover, scatter plots enable users to assess similarity between observations through distance, aiding in the detection of clusters, outliers, and degrees of class separation.

A scatter plot may be regarded as the quintessential graph for showing the relationship between two numerical variables. However, surprisingly perhaps, it has limitations when it comes to illustrating the \textit{statistical dependence} between variables. In particular, it includes information about the marginal distributions of the variables, which is irrelevant to their dependence. This excess information can obscure the true relationship between the variables. To address this, we turn our focus to rank plots, i.e., scatter plots of data ranks, which omit information related to marginal distributions and rest on solid mathematical theory.

The main contributions of this paper are: (1) explaining, through copula function theory, the inadequacies of scatter plots for dependence visualization, and the preference for measures like Spearman's rank correlation over Pearson's correlation; (2) advocating for rank plots as more suitable tools for analyzing statistical dependence; (3) providing interpretation guidelines for common patterns in rank plots; (4) introducing a novel graphical ensemble that we call a \textit{dplot,} for performing a comprehensive analysis of the relationship between two continuous random variables; and (5) all the Julia programming code and data sets used for calculations and generating figures is available for reproducibility.

The remainder of this paper is structured as follows: A review of relevant literature is presented in \cref{sec:relatedwork}, followed by a discussion on how marginal distributions influence scatter plots under an identical dependence relationship in \cref{sec:marginals}. In \cref{sec:copulas} we briefly delve into copula function theory, while in \ref{sec:dependence} we revisit key concepts of dependence, association measures, and graphical representations. In \cref{sec:dplot} we introduce the concept of a dplot, and in \cref{sec:realdata} we show their utility through several examples employing real data. Finally, \cref{sec:conclusions} contains the main conclusions and a discussion, and \cref{sec:reproduce} provides a link to the Julia programming code for reproducibility of all calculations and figures.

\section{Related work}
\label{sec:relatedwork}

Statistical dependence is often quantified through computational methods, yet the value of visualization techniques should not be underestimated, as illustrated by Anscombe's quartet~\cite{Anscombe73} and other synthetic data sets (see \cite{Matejka17}).

Scatter plots are the predominant method for visualizing the relationship between continuous random variables, with scatter plot matrices (SPLOM) extending this approach to pairwise analysis of several variables. For the latter, correlation matrices or \textit{corrgrams}~\cite{Friendly02}, which represent correlation measures through a color coding, are frequently employed. Although Pearson's correlation coefficient is the default measure in many software packages, it is not always the most appropriate choice, particularly if the data does not follow a joint Gaussian distribution. We discuss preferable alternatives in \cref{subsec:depcon}.

Several research works have been conducted on the perception of Pearson's correlation coefficient in scatter plots. Pioneering work by Doherty et al.~\cite{Doherty2007} focused on absolute estimates, while other posterior works have examined discriminative judgments by analyzing \textit{just-noticeable differences} (JND) between the correlations of two scatter plots presented simultaneously. Rensink and Baldridge~\cite{Rensink2010}, as well as Harrison et al.~\cite{Harrison2014}, have found that correlation adheres to Weber's Law, suggesting a linear relationship between correlation and JND. Conversely, Kay and Heer~\cite{Kay16} proposed a log-linear model for this relationship. A critical limitation of these studies is their reliance on Gaussian-distributed data or data generated through linear regression models~\cite{Doherty2007}. In contrast, Sher et al.~\cite{Sher2017} expanded the data set variety by altering aspects like density, shape, and number of clusters, ultimately questioning the reliability of human estimates of Pearson correlation in diverse scatter plots and challenging their utility. Recently, Strain et al.~\cite{Strain23a,Strain23b} have studied the perception of correlation when varying aspects such as contrast and point size, proposing solutions to mitigate underestimating correlation judgements. The rank plots under study in this work could be used to visually assess the degree of association between continuous random variables. However, note that a complementary user study falls beyond the scope of this paper.

Wilkinson et al.~\cite{Wilkinson05} developed a collection of graph-theoretic \textit{scagnostics}, which are measures related to scatter plots designed to aid in the exploration of large scatter plot matrices. Specifically, they considered the squared Spearman's correlation coefficient to measure monotonicity, i.e., trends or degree of association (see~\cite{Dang13}). In this paper, we employ rank plots to visualize monotonic relationships, which offer greater insight into dependence than a single numerical value. Moreover, we use Spearman's correlation coefficient (not squared) along with Schweizer-Wolff's dependence measure~\cite{Schweizer1981} due to their stronger theoretical foundation and advantages.

This paper relies on copula function theory (to explore dependence between random variables), which has not been fully exploited in the visualization literature. Previous works by Hazarika et al.~\cite{Hazarika18,Hazarika19} applied copulas to visualize uncertainty and to analyze large-scale multivariate simulation data. However, our focus is on visualizations that specifically facilitate the assessment and interpretation of dependence through copulas and their transformations.

Lastly, while techniques exist for indicating independence in categorical data, such as mosaic plots~\cite{Friendly01}, our study concentrates solely on continuous random variables.

\section{Effect of marginal distributions on scatter plots}
\label{sec:marginals}


Basic probability theory states that the relationship between two random variables is encapsulated within their joint probability density or cumulative distribution functions, whereas marginals alone lack such dependency information. However, scatter plots inherently reflect data from marginal distributions. Consequently, distinct scatter plots may exhibit notable differences even when their associated variables adhere to the same dependency structure.

\begin{figure}
  \centering
  \includegraphics[clip=true,width=\columnwidth]{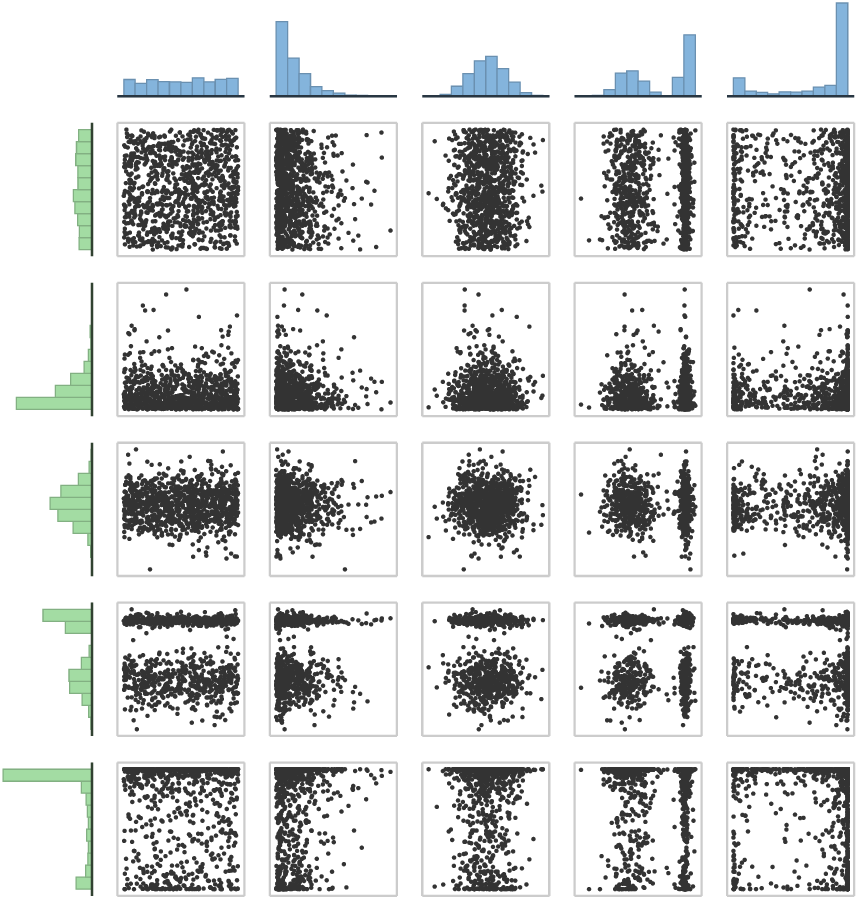} 
  \caption{Bivariate scatter plots of independent continuous random variables, but with different marginal distributions, as indicated by histograms. Specifically, we generated pairs of five marginal distributions in the following order: uniform, monotone, non-monotone unimodal, bimodal, and ``skewed'' bimodal.}
  \label{fig:indeplots_new}
\end{figure}

\Cref{fig:indeplots_new} shows 25 different scatter plots for the simplest probabilistic relationship between two random variables: \textit{independence}. Despite sharing the same dependence structure (in this case, independence), the scatter plots are quite diverse, due to variations in the shapes of their corresponding marginal distributions. In this example we have selected five types of marginal distributions (see Galtung's classification of distributions~\cite{Galtung67}): (1) Uniform, (2) unimodal monotone (peak on the left), (3) unimodal non-monotone (peak in the middle), (4) bimodal, and (5) ``skewed'' bimodal.

This example suggests that traditional scatter plots may not be ideal for visualizing dependence, since they are not invariant under changes in marginal distributions. However, as we will elucidate in the subsequent sections, a unique type of scatter plot, known as a rank plot (i.e., a scatter plot of ranks), adequately represents dependence information while remaining invariant to marginal distributions. A key characteristic of rank plots is that their associated marginals are always uniform. In \cref{fig:indeplots_new} the scatter plot located in the top-left corner is precisely a rank plot. Note that the uniform distribution of the plotted points clearly indicates independence (roughly speaking, the values of one variable are unaffected by those of the other). Thus, in general, we can infer that traditional scatter plots distort the information contained in rank plots by incorporating details about marginal distributions, potentially leading to misinterpretations when assessing dependence. Subsequent sections will present the main copula theory results supporting the use of rank plots.


\section{Bivariate copulas}
\label{sec:copulas}

A key concept regarding statistical dependence between continuous random variables is \textit{copula functions}, which was introduced in 1959 by Sklar \cite{Sklar1959}. He proved that, for a given vector $(X,Y)$ of two continuous random variables, there exists a unique function $C_{X,Y}:[0,1]^2\rightarrow[0,1]$ such that:
\begin{equation}\label{eq:sklar}
    F_{X,Y}(x,y) = C_{X,Y}\left(F_X(x),F_Y(y)\right),
\end{equation}
where $F_{X,Y}(x,y)=P(X\leq x,Y\leq y)$ is the joint cumulative distribution function of $(X,Y),$ and $F_X(x)=P(X\leq x)$ and $F_Y(y)=P(Y\leq y)$ the marginal cumulative distribution functions of $X$ and $Y,$ respectively. $C_{X,Y}$ is called \textit{copula function} of the random vector $(X,Y)$ and represents the unique functional link between the joint distribution and its marginals. Since marginal distributions have no information about how each random variable interacts with others, all the information about the dependence between random variables is contained in their underlying copula. Thus, \textit{Sklar's Theorem} implies that any proposal to analyze, measure or visualize dependence should be based only on the information that can be obtained from the underlying copula of the random variables.

Basic results and properties about copula functions may be found in \cite{Nelsen2006} or \cite{Joe2015}. For example, the underlying copula of a random vector is invariant under continuous strictly increasing transformations of its random variables:
\begin{equation}\label{eq:scale}
    C_{X,Y} = C_{g(X),h(Y)}\,\qquad g\uparrow\,,\,h\uparrow.
\end{equation}
This last property implies that the random vector $(X,Y)$ has exactly the same dependence structure as $\left(g(X),h(Y)\right)$, even though they could have different marginals. Consequently, dependence measures and visualizations aimed at illustrating only dependence should be identical in both cases.

Furthermore, recall from basic probability theory that for any continuous random variable $X$ with strictly increasing cumulative probability distribution function $F_X$, the transformed random variable defined as $U=F_X(X)$ always has a continuous uniform distribution over the closed interval $[0,1].$ In addition, its cumulative distribution function is the identity function $F_U(u)=u$ for $0\leq u\leq 1.$ These facts, combined with \eqref{eq:sklar} and \eqref{eq:scale}, have several important implications:
\begin{itemize}
    \item[(C1)] For any vector $(U,V)$ of continuous Uniform$(0,1)$ random variables: $F_{U,V}(u,v)=C_{U,V}(u,v).$ In other words, copula functions may be regarded as joint distributions with Uniform$(0,1)$ marginals.
    \item[(C2)] The random vectors $(X,Y)$ and $\left(F_X(X),F_Y(Y)\right)$ have the same underlying copula function, due to \eqref{eq:scale}, and therefore exactly the same dependence relationship.
    \item[(C3)] Even though $(X,Y)$ and $\left(F_X(X),F_Y(Y)\right)$ share the same copula (i.e., dependence structure), their scatter plots may look considerably different, since the latter has Uniform$(0,1)$ marginals but $(X,Y)$ typically will not.
\end{itemize}

Hence, scatter plots do not have a unique representation for the same type of dependence. Also, as a consequence of (C2) and (C3), we may consider the scatter plot of $\left(F_X(X),F_Y(Y)\right),$ which has Uniform$(0,1)$ marginals, as a canonical dependence representation of any other random vector $(X,Y)$ with the same dependence relationship but with any other marginal distributions. Lastly, as a consequence of (C1), such canonical representation would be a scatter plot of a random vector with Uniform$(0,1)$ marginals and joint cumulative distribution equal to the underlying copula. In other words, the joint cumulative distribution of $\left(F_X(X),F_Y(Y)\right)$ would be the copula $C_{X,Y}.$ Thus, a scatter plot of observations from $\left(F_X(X),F_Y(Y)\right)$ is a valid way to represent the information of a copula.

For example, consider the dependence structure associated with independence. Recall that two random variables $X$ and $Y$ are independent if and only if their joint distribution is equal to the product of its marginals, that is $F_{X,Y}(x,y)=F_X(x)F_Y(y).$ Therefore, as a consequence of Sklar's theorem (\ref{eq:sklar}) their unique underlying copula is given by:
\begin{equation}\label{eq:productcopula}
        C_{X,Y}(u,v) = uv = \Pi(u,v),
\end{equation}
usually known as the independence or \textit{product copula}. For example, the independent data sets associated with the 25 scatter plots in \cref{fig:indeplots_new} all have the same underlying copula $\Pi$ (despite having different marginals). Moreover, the canonical dependence scatter plot in this case would be the one in the upper left corner, which is a continuous uniform distribution over the unit square, where the marginals are uniform.


\begin{figure}[t]
    \centering

  \begin{minipage}{0.50\columnwidth}
   \centering
    \begin{psfrags}

      \psfrag{u}[cc][cc]{\footnotesize $U$}
      \psfrag{v}[cc][cc]{\footnotesize $V$}
      \psfrag{P(u,v)}[cc][cc]{\footnotesize $\Pi(u,v) = uv$}

     \includegraphics[clip=true,height=3.4cm]{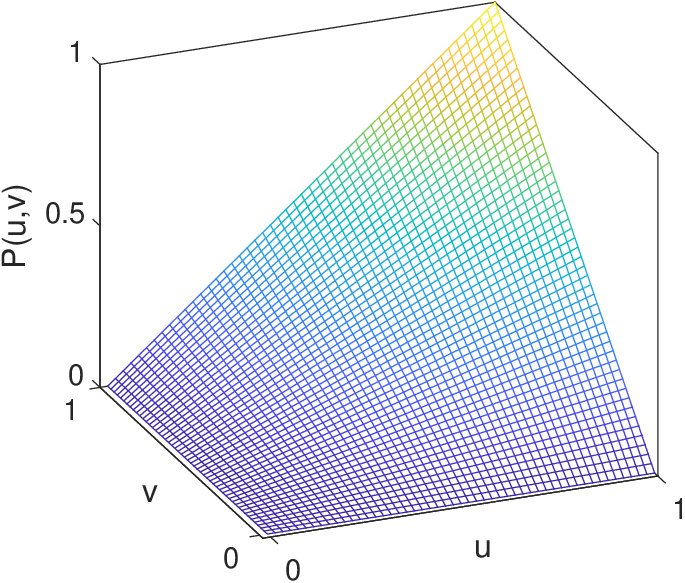}\\ \smallskip

    \end{psfrags}

   \hspace{0.38cm} \footnotesize (a) Product copula $\Pi$
  \end{minipage}
  \hfill
  \begin{minipage}{0.42\columnwidth}
   \centering

    \begin{psfrags}

      \psfrag{U}[cc][cc]{\footnotesize $U$}
      \psfrag{V}[cc][cc]{\footnotesize $V$}
      \includegraphics[clip=true,height=3.4cm]{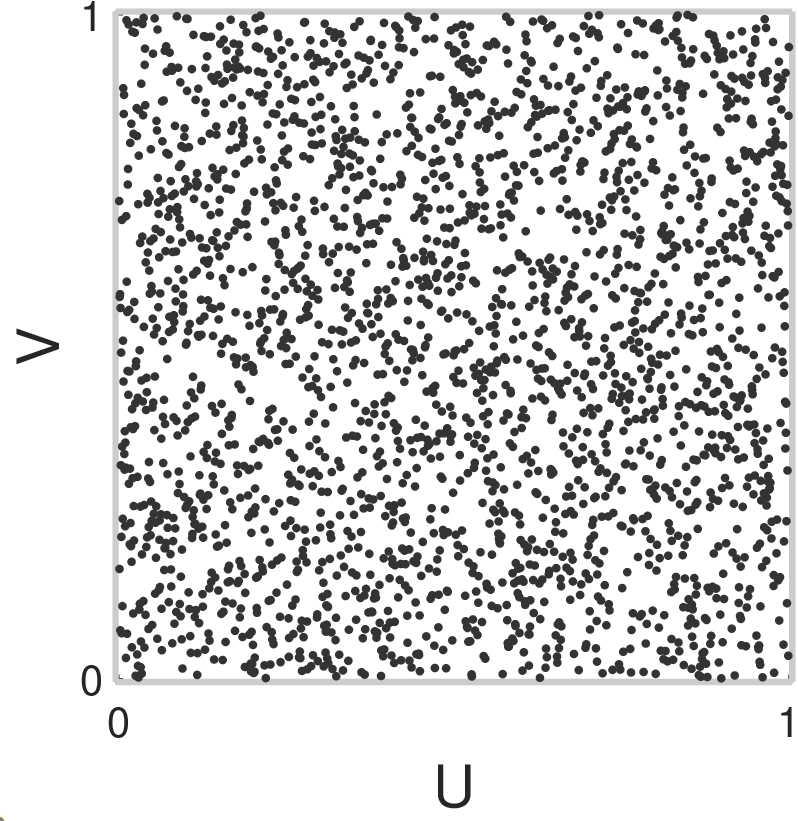}\\

    \end{psfrags}

   \hspace{0.38cm} \footnotesize (a') Observations from $\Pi$
  \end{minipage}

\mbox{} \\ \bigskip

  \begin{minipage}{0.50\columnwidth}
   \centering
    \begin{psfrags}

      \psfrag{u}[cc][cc]{\footnotesize $U$}
      \psfrag{v}[cc][cc]{\footnotesize $V$}
      \psfrag{M(u,v)}[cc][cc]{\footnotesize $M(u,v) = \min(u,v)$}

     \includegraphics[clip=true,height=3.4cm]{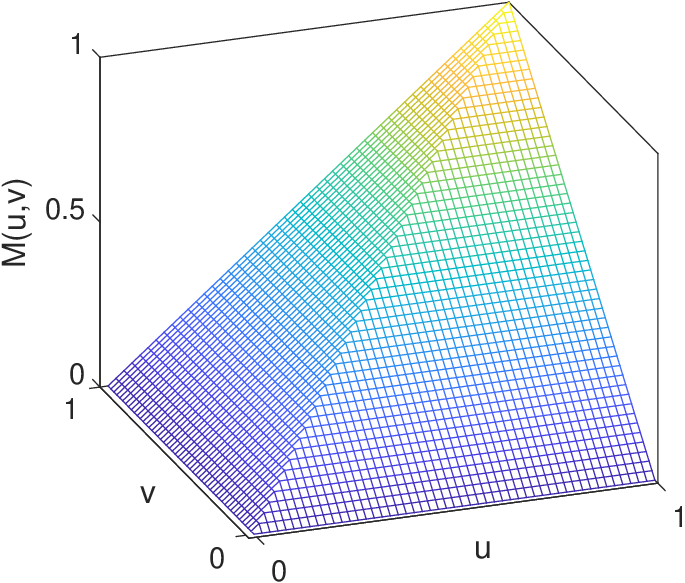}\\ \smallskip

    \end{psfrags}

   \hspace{0.38cm} \footnotesize (b) Upper bound copula $M$
  \end{minipage}
  \hfill
  \begin{minipage}{0.42\columnwidth}
   \centering

    \begin{psfrags}

      \psfrag{U}[cc][cc]{\footnotesize $U$}
      \psfrag{V}[cc][cc]{\footnotesize $V$}
     \includegraphics[clip=true,height=3.4cm]{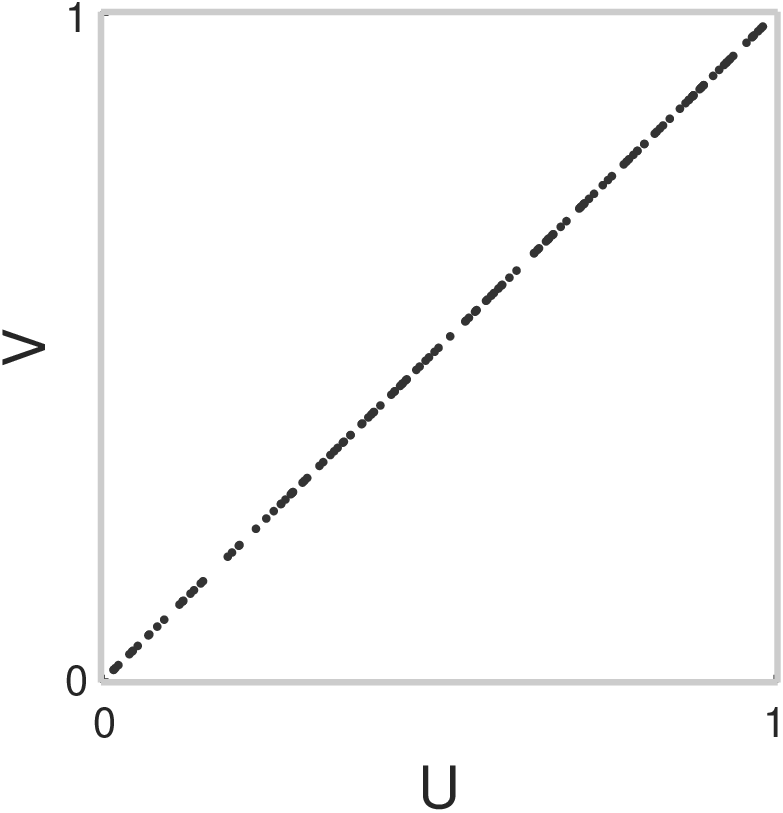}\\

    \end{psfrags}

   \hspace{0.38cm} \footnotesize (b') Observations from $M$
  \end{minipage}

\mbox{} \\ \bigskip

  \begin{minipage}{0.50\columnwidth}
   \centering
    \begin{psfrags}

      \psfrag{u}[cc][cc]{\footnotesize $U$}
      \psfrag{v}[cc][cc]{\footnotesize $V$}
      \psfrag{W(u,v)}[cc][cc]{\footnotesize \hspace{-0.5cm} $W(u,v) = \max(u+v-1,0)$}

     \includegraphics[clip=true,height=3.4cm]{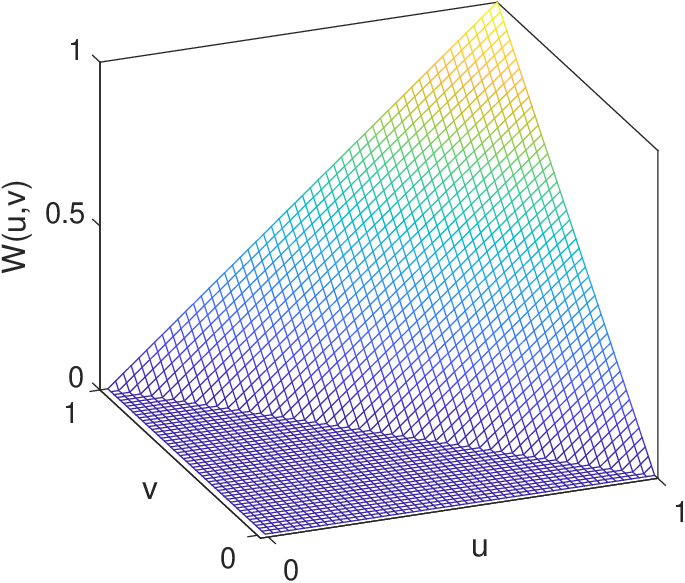}\\ \smallskip

    \end{psfrags}

   \hspace{0.38cm} \footnotesize (c) Lower bound copula $W$
  \end{minipage}
  \hfill
  \begin{minipage}{0.42\columnwidth}
   \centering

    \begin{psfrags}

      \psfrag{U}[cc][cc]{\footnotesize $U$}
      \psfrag{V}[cc][cc]{\footnotesize $V$}
     \includegraphics[clip=true,height=3.4cm]{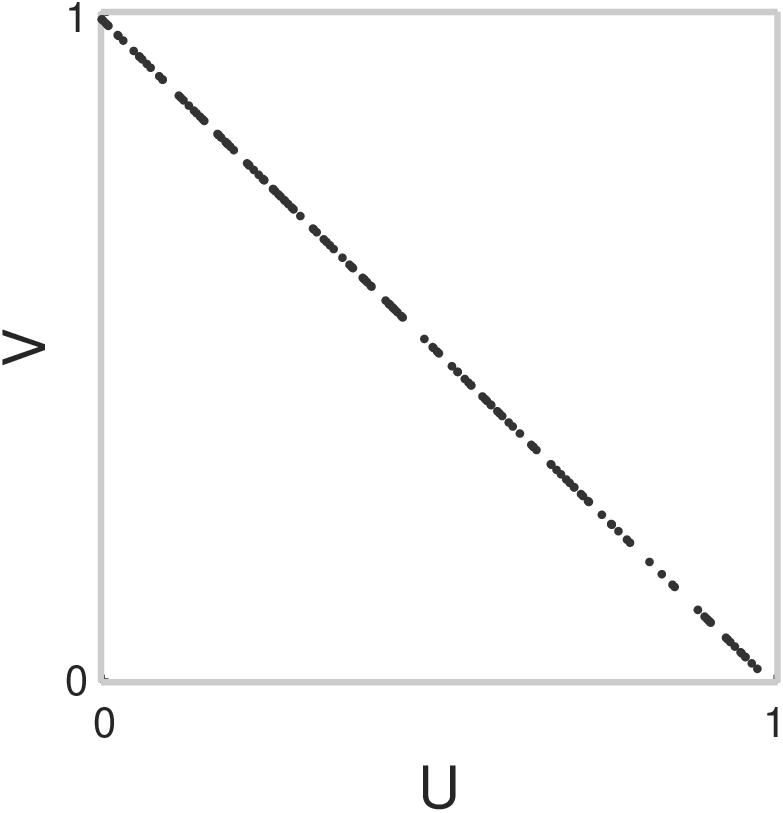}\\

    \end{psfrags}

   \hspace{0.38cm} \footnotesize (c') Observations from $W$
  \end{minipage}

  \caption{Basic copula functions $\Pi$ (indicating independence),  $M$ (illustrating a strictly increasing relationship), and $W$ (representing a strictly decreasing relationship), together with observations from these copulas.}
\label{fig:copulasbasicas}
\end{figure}



\section{Dependence types, measures, and plots}
\label{sec:dependence}

\subsection{Quadrant dependence} \label{subsec:qd}

According to the results from \cref{sec:copulas}, if observations from $\left(F_X(X),F_Y(Y)\right)$ appear uniformly distributed over the unit square we may conjecture that the random variables are independent (or exhibit a very weak dependence). Departures from this scenario imply some kind of dependence relationship that requires assessment. E. Lehmann described a comprehensive catalog of general types of dependencies~\cite{Lehmann1966}, while R. Nelsen identified them in terms of copulas~\cite{Nelsen2006}. The most general and simple type is known as \textit{quadrant dependence}, which may be positive (PQD) or negative (NQD):
\begin{eqnarray}\label{eq:PQD}
\text{PQD}(X,Y) &\Leftrightarrow& P(X\leq x,Y\leq y) \geq P(X\leq x)P(Y\leq y) \nonumber \\
                &\Leftrightarrow& P(X>x,Y>y) \geq P(X>x)P(Y>y) \nonumber \\
                &\Leftrightarrow& C_{X,Y}(u,v) \geq \Pi(u,v)=uv.
\end{eqnarray}
\begin{eqnarray}\label{eq:NQD}
\text{NQD}(X,Y) &\Leftrightarrow& P(X\leq x,Y>y) \geq P(X\leq x)P(Y>y) \nonumber \\
                &\Leftrightarrow& P(X>x,Y\leq y) \geq P(X>x)P(Y\leq y) \nonumber \\
                &\Leftrightarrow& C_{X,Y}(u,v) \leq \Pi(u,v)=uv.
\end{eqnarray}
Intuitively, $\text{PQD}(X,Y)$ implies that the joint probability that values from $X$ and $Y$ are simultaneously small (or simultaneously large) is greater than or equal to the analogous probability if the variables were independent. As a consequence of \eqref{eq:sklar}, the copula function of $X$ and $Y$ will be greater than or equal to the product copula. We can also interpret that small (large) values of $X$ tend to be more likely associated with small (large) values of $Y$. Conversely, $\text{NQD}(X,Y)$ implies that small (large) values of $X$ tend to be more likely associated to large (small) values of $Y$ (and the copula of $X$ and $Y$ will be less than or equal to $\Pi$). Thus, roughly speaking, PQD and NQD are associated with increasing and decreasing trends, respectively.

Regression models that fall into the category of quadrant dependence are those of the form $Y=\varphi(X)+\varepsilon$, where $\varphi$ is a continuous and strictly monotone function (increasing or decreasing), and $\varepsilon$ is a random noise variable centered around zero. Particularly, if $\varphi$ is a linear function we have the case of linear regression. In terms of observations from $\left(F_X(X),F_Y(Y)\right),$ PQD can be identified when they appear close to the graph of $v=u,$ for $0\leq u\leq 1$ (i.e., the ``main'' diagonal). Alternatively, for NQD the points will lie close to the ``secondary'' diagonal corresponding to $v=1-u$.


Another significant aspect concerning copulas is the ability to ascertain the proximity to PQD, NQD and independence. As we have seen, a uniform distribution on $[0,1]^2$ of observations from $\left(F_X(X),F_Y(Y)\right)$ indicates independence. Therefore, a departure from uniformity indicates some type of dependence. The maximum deviation from independence can be established by applying Sklar's theorem \eqref{eq:sklar} to the Fr\'echet-Hoeffding bounds \cite{Hoeffding1940,Frechet1951} for bivariate joint distributions:
\begin{equation}\label{eq:FHbounds}
    W(u,v) \leq C_{X,Y}(u,v) \leq M(u,v),
\end{equation}
where $W(u,v)=\max\{u+v-1,0\}$ and $M(u,v)=\min\{u,v\}$ are also copulas, see \cref{fig:copulasbasicas}. It is straightforward to prove that if $Y=\varphi(X)$, where $\varphi$ is a continuous strictly increasing function, then $C_{X,Y}=M$, indicating PQD. In such instances, observations from $\left(F_X(X),F_Y(Y)\right)$ align precisely along the main diagonal $v=u$. Conversely, if $\varphi$ is a continuous strictly decreasing function then $C_{X,Y}=W$, signifying NQD. In that case, the scatter plot of observations from $\left(F_X(X),F_Y(Y)\right)$ would exclusively contain points along the secondary diagonal $v = 1-u$ (see \cref{fig:copulasbasicas}).

For a general regression model $Y=\varphi(X)+\varepsilon$ with $\varphi$ continuous and strictly monotone, the smaller the variability of $\varepsilon$ the closer the underlying copula gets to one of the Fr\'echet-Hoeffding bounds (\ref{eq:FHbounds}), and the closer a scatter plot of observations from $\left(F_X(X),F_Y(Y)\right)$ gets to one of the diagonal lines $v=u$ or $v=1-u$. Furthermore, as the variability of $\varepsilon$ increases the underlying copula will increasingly resemble the independence copula $\Pi$, and observations from $\left(F_X(X),F_Y(Y)\right)$ will more closely resemble a uniform distribution across the unit square.

\subsection{Dependence versus concordance}
\label{subsec:depcon}


According to Sklar's theorem \eqref{eq:sklar}, since all the information about the dependence relationship between a pair of continuous random variables $(X,Y)$ is in their underlying unique copula $C_{X,Y}$, any attempt to measure dependence must be copula based only, considering how far $C_{X,Y}$ is from the independence copula $\Pi$. Schweizer and Wolff (see \cite{Schweizer1981}) proposed a measure $\sigma_{\,X,Y}$ based on the $L_1$ distance between the graphs of $C_{X,Y}$ and $\Pi$ defined as follows:
\begin{equation}\label{eq:schweizer}
    \sigma_{\,X,Y} = 12\!\!\int_{0}^{\,1}\!\!\!\!\!\int_{0}^{\,1} \left|C_{X,Y}(u,v)-uv\right|dudv.
\end{equation}
It can be shown that $\sigma_{\,X,Y}$ satisfies the properties of a \textit{measure of dependence}, as defined in \cite{Nelsen2006}. In particular, the double integral is multiplied by $12$ in order to provide a normalized measure between $0$ and $1$. Note that the farthest a copula can be from $\Pi$ is one of the Fr\'echet-Hoeffding bounds \eqref{eq:FHbounds}, and the $L_1$ distance between $M$ and $\Pi$, and between $\Pi$ and $W$, is $1/12$.

Furthermore, observe that $\sigma_{\,X,Y}$ only depends on the copula $C_{X,Y}$ (e.g., it does not depend on marginal distributions). Also, $\sigma_{\,X,Y}=0$ if and only if $C_{X,Y}=\Pi$, which occurs if and only if $X$ and $Y$ are independent. Note that this desirable unique characterization of independence is not provided by the popular Pearson's correlation coefficient, for which a zero value does not necessarily imply independence.



Embrechts et al.~\cite{Embrechts1999} analyzed in detail additional pitfalls of Pearson's correlation coefficient $r_{\,X,Y}$ in terms of measuring dependence. If $Y=g(X)$ with $g$ strictly monotone, then in general $r_{\,X,Y}$ fails to achieve any of its extreme values $\{-1,+1\}$, for example, if $X$ is Uniform$(0,1)$ and $Y=X^t$ for any $t>1.$ In general $r_{\,X,Y}\neq r_{\log X,\log Y}$ even though the copula of $(X,Y)$ and $(\log X,\log Y)$ is exactly the same, as a consequence of \eqref{eq:scale}. Moreover, $r_{\,X,Y}$ does not exist for every pair of random variables since it depends on the existence of the marginal variances. Thus, it is not even a general linearity measure, for example, if $X$ is Cauchy distributed and $Y=a+bX$, then even though there is a clear linear relationship between $X$ and $Y$, $r_{\,X,Y}$ does not even exist.

The pitfalls of Pearson's correlation have their root in its link to marginal characteristics that have no information about dependence of random variables, since:
\begin{equation}\label{eq:pearson}
    r_{\,X,Y} = \frac{1}{\sqrt{V(X)V(Y)}}\int_{0}^{\,1}\!\!\!\!\!\int_{0}^{\,1} \left[C_{X,Y}(u,v)-uv\right]dF_X^{-1}(u)dF_Y^{-1}(v).
\end{equation}
Pearson's correlation coefficient mixes information from the dependence (given by the underlying copula $C_{X,Y}$) with marginal information that has nothing to do with dependence.

We could improve this correlation coefficient making it marginal-free by considering $\rho_{\,X,Y} = r_{F_X(X),F_Y(Y)}$. Since $F_X(X)$ and $F_Y(Y)$ are continuous Uniform$(0,1)$ distributions, their variances always exist and are equal to $\frac{1}{12}$. Furthermore, the dependence information from the random vector $\left(F_X(X),F_Y(Y)\right)$ is the same as for $(X,Y)$, as a consequence of (\ref{eq:scale}). In fact, $\rho_{\,X,Y}$ is Spearman's correlation \cite{Spearman1904}, also known as Spearman's \textit{concordance measure,} and \eqref{eq:pearson} becomes:
\begin{equation}\label{eq:spearman}
    \rho_{\,X,Y} = 12\!\!\int_{0}^{\,1}\!\!\!\!\!\int_{0}^{\,1} \left[C_{X,Y}(u,v)-uv\right]dudv.
\end{equation}

It can be shown that $\rho_{\,X,Y}$ satisfies the properties of a \textit{measure of concordance}, as defined in \cite{Nelsen2006}. The only difference between $\sigma_{\,X,Y}$ in \eqref{eq:schweizer} and $\rho_{\,X,Y}$ is that the integrand in \eqref{eq:spearman} is not in absolute value. Therefore $\rho_{\,X,Y}=0$ does not necessarily imply independence, and it cannot be considered a dependence measure.


From (\ref{eq:PQD}), (\ref{eq:NQD}), (\ref{eq:schweizer}), and (\ref{eq:spearman}) we have the following relationships:
\begin{itemize}
    \item[(B1)] $|\rho_{\,X,Y}| \leq \sigma_{\,X,Y}\,;$
    \item[(B2)] $\sigma_{\,X,Y}=0$ implies $\rho_{\,X,Y}=0$, but not vice versa;
    \item[(B3)] $\text{PQD}(X,Y)$ if and only if $\sigma_{\,X,Y} = \rho_{\,X,Y}\,;$
    \item[(B4)] $\text{NQD}(X,Y)$ if and only if $\sigma_{\,X,Y} = -\rho_{\,X,Y}\,;$
    \item[(B5)] $X$ and $Y$ are neither PQD nor NQD if and only if $|\rho_{\,X,Y}|<\sigma_{\,X,Y}\,.$
\end{itemize}

The pair of values $(\rho_{\,X,Y},\sigma_{\,X,Y})$ provides valuable insights into the relationship between two continuous random variables. By understanding these two values, we can promptly discern their independence and, if dependent, classify whether the relationship is PQD, NQD, or neither of these.

\subsection{Non quadrant dependence}\label{subsec:nonqd}

If $|\rho_{\,X,Y}|<\sigma_{\,X,Y}$ then $X$ and $Y$ are not quadrant dependent. This implies, as inferred from equations \eqref{eq:PQD} and \eqref{eq:NQD}, that within certain subsets of the unit square $C_{X,Y}$ exceeds $\Pi$, while in complementary regions $C_{X,Y}$ falls below $\Pi$.  Equivalently, within certain areas of the support of $(X,Y)$ the relationship between $X$ and $Y$ is PQD, while in other regions it is NQD. In subsequent subsections we analyze two primary approaches for achieving this behavior (other alternatives exist).

\subsubsection{Convex linear combinations}\label{subsubsec:convex}

If $C_1$ and $C_2$ are two copulas, and $0\leq\theta\leq 1$, then as proved in \cite{Nelsen2006}, any convex linear combination of them is also a copula:
\begin{equation}\label{eq:convex}
    C_{1,2,\theta}(u,v) = (1-\theta)C_1(u,v) + \theta C_2(u,v).
\end{equation}
If for example we choose as $C_1$ a PQD copula and as $C_2$ a NQD one, then \eqref{eq:convex} would be a non quadrant dependence copula, and by \eqref{eq:sklar} we can build a joint cumulative distribution function for a vector $(X,Y)$ of continuous random variables with such copula and any given continuous marginals, see for example case R4 in \cref{fig:rankplots}.


\subsubsection{Gluing copulas}\label{subsubsec:gluing}

It is possible to combine two copulas through a copula construction technique known as ``gluing copulas''  \cite{Siburg2008}. Given two copulas $C_1$ and $C_2,$ and a fixed value $0<\theta<1,$ called a \textit{gluing point,} we  scale and \textit{glue} them (intuitively, ``concatenate'' them) horizontally according to a vertical partition of the unit square. In particular, $C_1$ is scaled to $[0,\theta]\times[0,1]$ and $C_2$ to $[\theta,1]\times[0,1].$ Finally, they are \textit{glued} into a single copula:
\begin{equation}\label{eq:gluing}
    C_{1,2,\theta}(u,v) = \begin{cases}
        \,\theta C_1\left(\frac{u}{\theta},v\right), \text{ if } 0\leq u\leq \theta, \\
        \,(1-\theta)C_2\left(\frac{u-\theta}{1-\theta},v\right)+\theta v, \text{ if } \theta\leq u\leq 1.
    \end{cases}
\end{equation}
If we choose as $C_1$ a PQD copula, and as $C_2$ a NQD copula, then \eqref{eq:gluing} would be a non quadrant dependence copula. By \eqref{eq:sklar} we can build a joint cumulative distribution function for a vector $(X,Y)$ of continuous random variables by $F_{X,Y}(x,y)=C_{1,2,\theta}\left(F_X(x),F_Y(y)\right)$, where $\text{PQD}(X,Y)$ if $X\leq F_X^{-1}(\theta)$ but $\text{NQD}(X,Y)$ if $X\geq F_X^{-1}(\theta)$ (see case R7 \cref{fig:rankplots}. The gluing copula technique is particularly useful for \textit{piecewise regression}~\cite{Erdely2017a}.


\subsection{Rank plots and empirical estimation}\label{subsec:empirical}

In practice, we usually have a random sample of paired observed values $\{(x_1,y_1),\ldots,(x_n,y_n)\}$ from a random vector $(X,Y)$ with an unknown joint probability distribution and also unknown marginal distributions. In such case, a natural replacement for observed values of $\left(F_X(X),F_Y(Y)\right)$ would be
$\{(u_k,v_k)=\left(F_n(x_k),G_n(y_k)\right):k=1,\ldots,n\}$ where $F_n$ and $G_n$ are unbiased and consistent estimations of the marginal distributions $F_X$ and $F_Y,$ respectively, known as empirical distribution functions \cite{Wasserman2006}. For continuous random variables there are no repeated values and therefore $nF_n(x_k)=\text{rank}(x_k),$ that is, the total number of observations from $X$ that are equal to or less than $x_k.$ Thus, a scatter plot of $(u_k,v_k)$ values is a bivariate plot of observed ranks scaled to lie in the unit square, which we will call a \textit{rank plot} (some authors call it a plot of \textit{pseudo-observations}~\cite{Hofert2018}), and constitutes an empirical approximation of a scatter plot from the unknown underlying copula function.

\begin{figure}[t]
  \centering

    \begin{minipage}{0.30\columnwidth}
   \centering
     \includegraphics[clip=true,width=\textwidth]{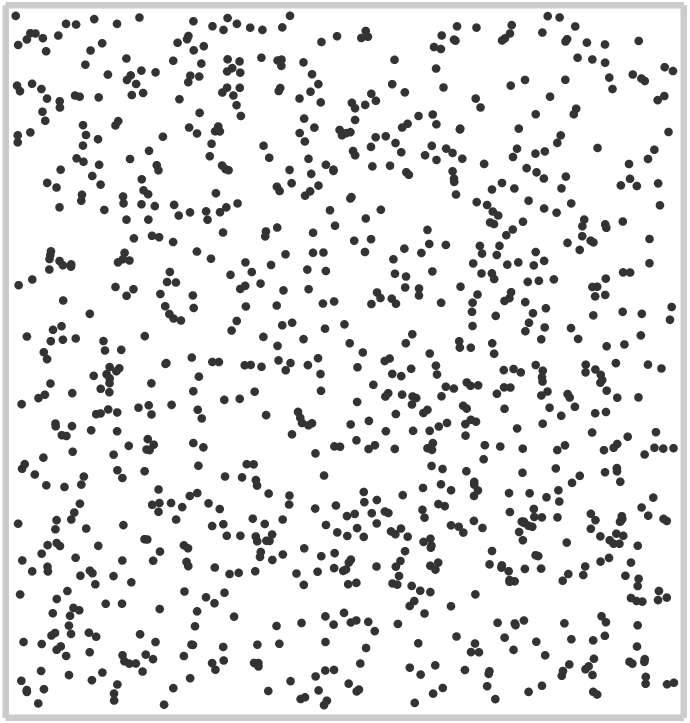}\\
     \footnotesize{R1} \\
  \end{minipage}
  \hfill
  \begin{minipage}{0.30\columnwidth}
   \centering
     \includegraphics[clip=true,width=\textwidth]{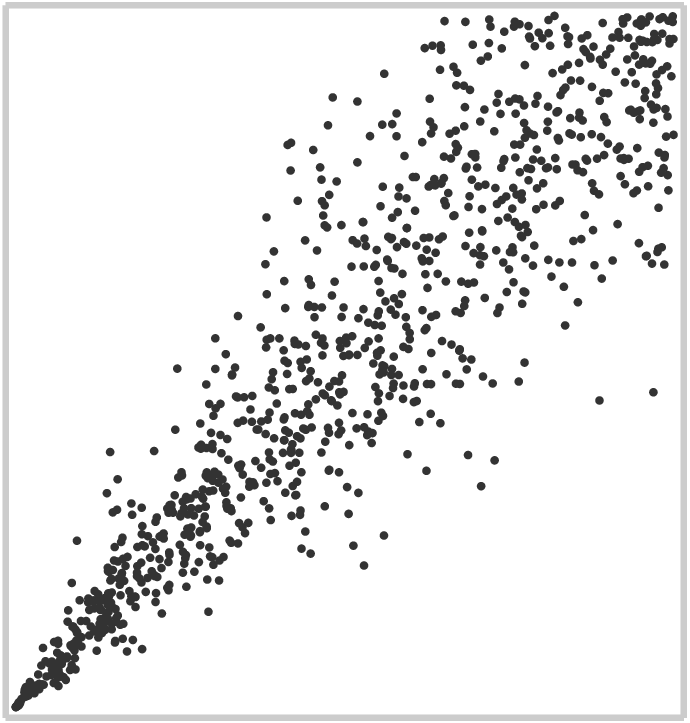}\\
     \footnotesize{R2} \\
  \end{minipage}
  \hfill
  \begin{minipage}{0.30\columnwidth}
   \centering
     \includegraphics[clip=true,width=\textwidth]{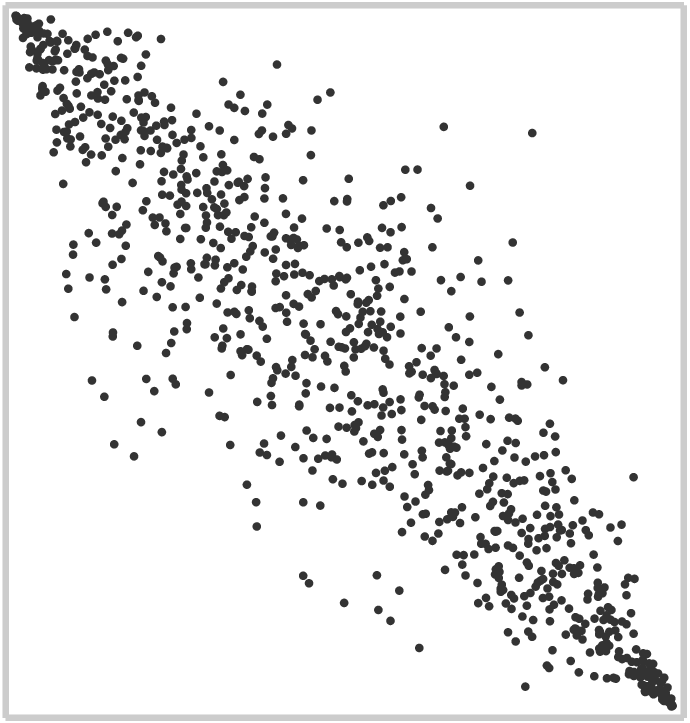}\\
     \footnotesize{R3} \\
  \end{minipage}

\mbox{} \\ \medskip

    \begin{minipage}{0.30\columnwidth}
   \centering
     \includegraphics[clip=true,width=\textwidth]{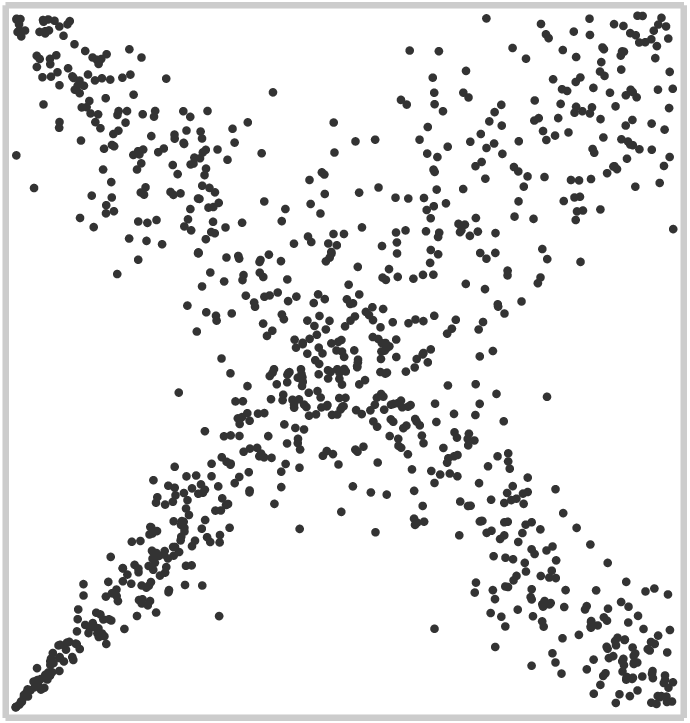}\\
     \footnotesize{R4} \\
  \end{minipage}
  \hfill
  \begin{minipage}{0.30\columnwidth}
   \centering
     \includegraphics[clip=true,width=\textwidth]{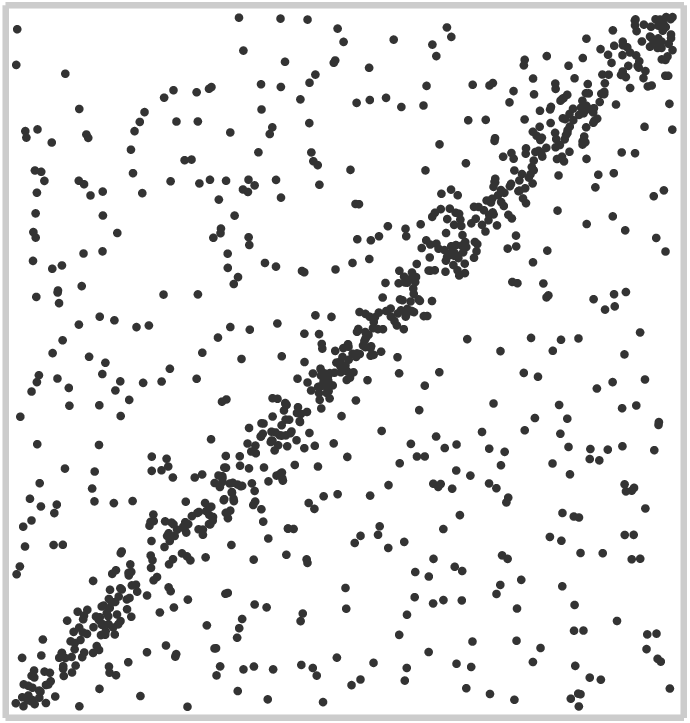}\\
     \footnotesize{R5} \\
  \end{minipage}
  \hfill
  \begin{minipage}{0.30\columnwidth}
   \centering
     \includegraphics[clip=true,width=\textwidth]{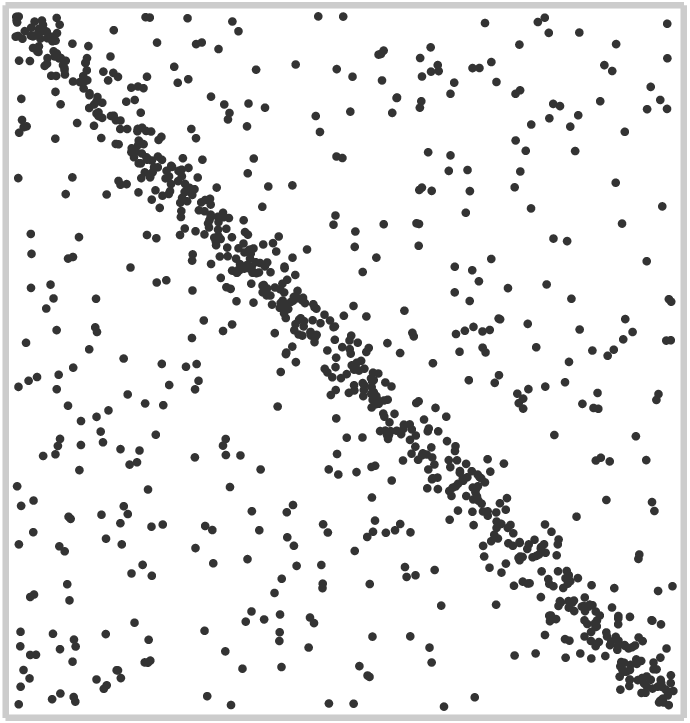}\\
     \footnotesize{R6} \\
  \end{minipage}

\mbox{} \\ \medskip

    \begin{minipage}{0.30\columnwidth}
   \centering
     \includegraphics[clip=true,width=\textwidth]{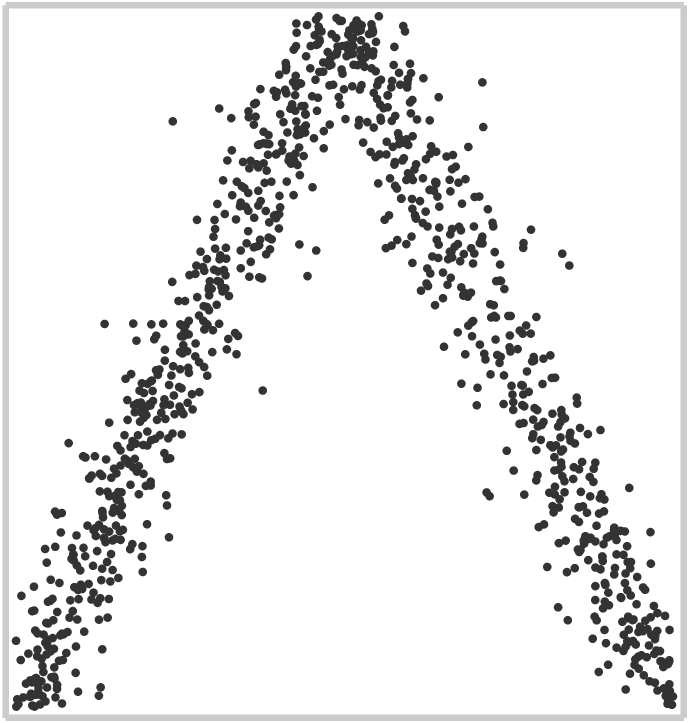}\\
     \footnotesize{R7} \\
  \end{minipage}
  \hfill
  \begin{minipage}{0.30\columnwidth}
   \centering
     \includegraphics[clip=true,width=\textwidth]{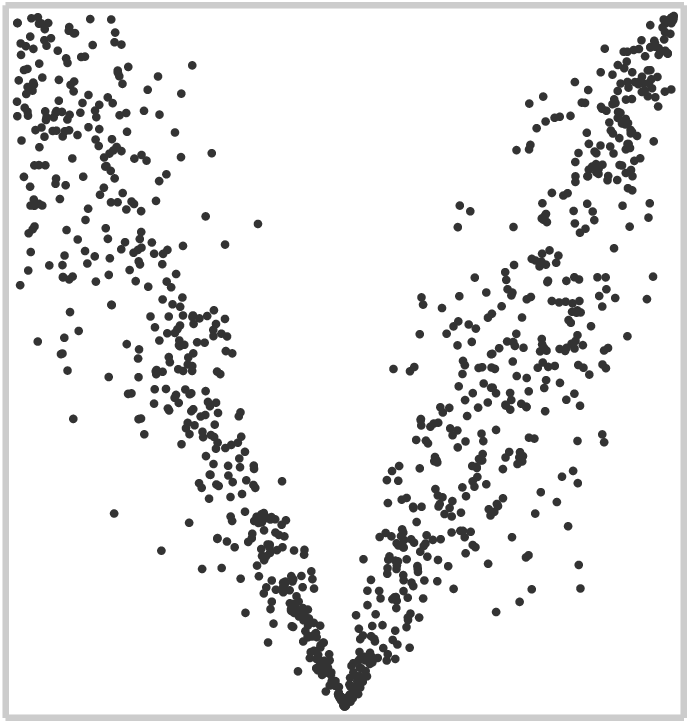}\\
     \footnotesize{R8} \\
  \end{minipage}
  \hfill
  \begin{minipage}{0.30\columnwidth}
   \centering
     \includegraphics[clip=true,width=\textwidth]{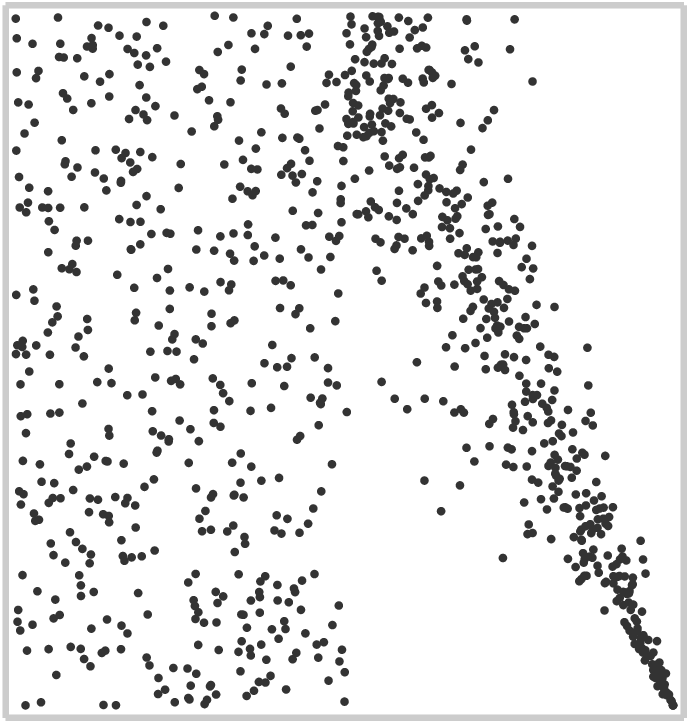}\\
     \footnotesize{R9} \\
  \end{minipage}

  \caption{Examples of rank plots associated with the guideline cases in \cref{tab:deptypes}. R1 (independence), R2 (PQD), and R3 (NQD). Convex linear combinations: R4 (PQD + NQD), R5 (independence + PQD), and R6 (independence + NQD). Gluing copulas: R7 (PQD | NQD), R8 (NQD | PQD), and R9 (independence | NQD).}
  \label{fig:rankplots}
\end{figure}

According to what has been already discussed in the previous sections, rank plots appropriately illustrate the dependence between the data variables. \Cref{tab:deptypes} and \cref{fig:rankplots} show a categorization of dependence types that can be used as a set of guidelines for interpreting rank plots. In practice, the described patterns might be quite clear, but sometimes not. To help in this empirical dependence assessment, in addition to rank plots we propose analyzing empirical estimations for Schweizer-Wolff's dependence measure \eqref{eq:schweizer} and Spearman's concordance \eqref{eq:spearman} to take advantage of their combined interpretation, as explained in subsection \ref{subsec:depcon}.

\begin{table}[tb]
  \caption{%
  	Guidelines for dependence assessment according to rank plot appearance.%
  }
  \label{tab:deptypes}
  \scriptsize%
  \centering%
  \begin{tabu}{%
  	  r%
  	  	*{7}{c}%
  	  	*{2}{r}%
  	}
  	\toprule
  	\textbf{Case} & \textbf{Rank plot appearance} & \textbf{Dependence assessment} \\
  	\toprule
  	R1 & Uniform & (close to) Independence \\
        \midrule
        R2 & Most points close to $v=u$ & PQD \\
        \midrule
        R3 & Most points close to $v=1-u$ & NQD \\
  	\midrule
        R4 & Some points close to $v=u$ & Convex linear combination \\
        { } & but some close to $v=1-u$  & of PQD and NQD \\
        \midrule
        R5 & Some points close to $v=u$ & Convex linear combination \\
        { } & but some uniformly distributed  & of PQD and independence \\
        \midrule
        R6 & Some points close to $v=1-u$ & Convex linear combination \\
        { } & but some uniformly distributed  & of NQD and independence \\
        \midrule
        R7 & PQD for $u\leq\theta$ & Gluing of PQD and NQD \\
        { } & and NQD for $u>\theta$ &  (in that order) \\
        \midrule
        R8 & NQD for $u\leq\theta$ & Gluing of NQD and PQD \\
        { } & and PQD for $u>\theta$ &  (in that order) \\
        \midrule
        R9 & Independence for $u\leq\theta$ & Gluing of independence \\
        { } & and NQD (or PQD) for $u>\theta$ &  and NQD (or PQD) \\
  	\bottomrule
  \end{tabu}%
\end{table}

The empirical estimation of the underlying bivariate copula is given by a function $C_n$, with domain the grid $\{0,\frac{1}{n},\ldots,\frac{n-1}{n}, 1\}^2$, defined as:
\begin{equation}\label{eq:copem}
    C_n\left(\frac{i}{n}\,,\,\frac{j}{n}\right) = \frac{1}{n}\sum_{k\,=\,1}^n \mathbf{1}_{\{x_k\,\leq\,x_{(i)}\,,\,y_k\,\leq\,y_{(j)}\}},
\end{equation}
where $x_{(i)}$ denotes the $i$-th order statistic, and $C_n(i/n,0 ) = 0 = C_n(0,j/n)$. The function $C_n$ is usually referred to as the \textit{empirical copula} though originally it was introduced as \textit{empirical dependence function} by \cite{Deheuvels1979}. In addition, the empirical estimation of \eqref{eq:spearman} is~\cite{Nelsen2006}:
\begin{equation}\label{eq:empspearman}
    \rho_n = \frac{12}{n^2-1}\sum_{i\,=\,1}^{n}\sum_{j\,=\,1}^{n} \left[C_n\left(\frac{i}{n}\,,\,\frac{j}{n}\right) - \frac{ij}{n^2}\right],
\end{equation}
while for \eqref{eq:schweizer} its empirical estimation $\sigma_n$ is obtained by replacing the differences in the sums in \eqref{eq:empspearman} with the absolute value of the differences.

\subsection{Diagonal sections}
\label{subsubsec:diagonals}

\begin{figure}[t]
  \centering
  \includegraphics[width=\columnwidth]{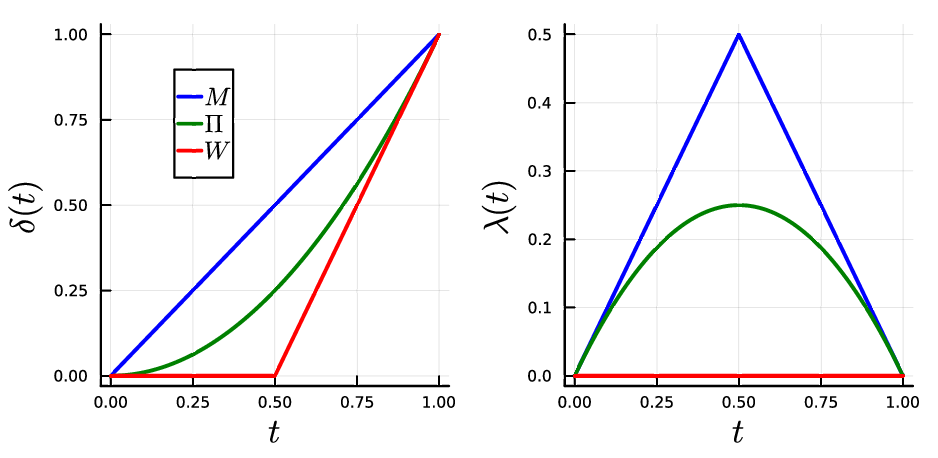}
  \caption{Main diagonal sections $\delta(t)=C(t,t)$ and secondary diagonal sections $\lambda(t)=C(t,1-t)$ for the Fr\'echet-Hoeffding bounds copulas $W$ and $M,$ and the product copula $\Pi.$}
  \label{fig:diagonals}
\end{figure}


Besides $\rho_n$ and $\sigma_n$, there are other characteristics of the copula that help in interpreting a rank plot. For example, it is useful to determine whether the copula is above or below $\Pi$, or how close it is to $M$ or $W$. Even though we could visualize $C_n$ either by a 3D surface plot (as in \cref{fig:copulasbasicas}) or a contour plot, it would be difficult to grasp the nuances that allow us to interpret the dependence structure in these visualizations. Alternatively, we propose visualizing the graphs of the main and secondary diagonal sections of the copula. These simplifications can be useful for detecting departures from independence, PQD, NQD, and especially to identify cases exhibiting both PQD and NQD, which may be analyzed through gluing copulas (see \cite{Erdely2017a,Erdely2010}).

The main diagonal section of a copula $C$ is given by $\delta_{\,C}(t)=C(t,t),$ and the secondary diagonal by $\lambda_{\,C}(t)=C(t,1-t).$ As an immediate consequence of \eqref{eq:FHbounds} we have:
\begin{equation}\label{eq:diag}
    \delta_W(t) = \max\{2t-1,0\} \leq \delta_{\,C}(t) \leq t = \delta_M(t),
\end{equation}
and
\begin{equation}\label{eq:diagsec}
    \lambda_W(t) = 0 \leq \lambda_{\,C}(t) \leq \min\{t,1-t\} = \lambda_M(t),
\end{equation}
for $0\leq t\leq 1$, where $\delta_W(t)$ and $\lambda_W(t)$ represent the main and secondary diagonal sections of $W$, respectively. Similarly, $\delta_M(t)$ and $\lambda_M(t)$ are diagonal sections of $M$. \Cref{fig:diagonals} shows these bounds, together with the diagonal sections from the independence copula, which are $\delta_{\,\Pi}(t)=t^2$ and $\lambda_{\,\Pi}(t)=t(1-t)$. In cases of PQD$(X,Y)$ we have $\delta_{\,X,Y}(t)\geq t^2$ and $\lambda_{\,X,Y}(t)\geq t(1-t).$ 
Likewise, for NQD$(X,Y)$ we have $\delta_{\,X,Y}(t)\leq t^2$ and $\lambda_{\,X,Y}(t)\leq t(1-t).$ 
If there is a crossing between $\delta_{\,\Pi}$ and $\delta_{\,X,Y}$ or $\lambda_{\,X,Y}$ then there would not be PQD or NQD, and the crossing point $t=\theta$ could be a gluing point of two copulas, as described in \eqref{eq:gluing}. 


\section{dplots}
\label{sec:dplot}


In order to comprehensively analyze bivariate data and effectively visualize the dependency structure between the variables (as discussed in the previous sections), along with characteristics depicted in traditional scatter plots, we advocate for the simultaneous visualization of:
\begin{itemize}
    \item[a)] The regular \textbf{scatter plot}, to visualize characteristics such as concrete data values, clusters, or outliers.
    \item[b)] The \textbf{rank plot}, to analyze the dependence between the variables without the distortion of the marginal distributions, and to identify it according to the categorization in \cref{tab:deptypes} and \cref{fig:rankplots}.
    \item[c)] Marginal \textbf{histograms}, to visualize marginal behavior of the probability density functions (pdfs), in cases where it could be hard to extract from the scatter plot (e.g., due to occlusion caused by a large number of displayed points),
    and understand their combined influence in the shape of the scatter plot (as in \cref{fig:indeplots_new}). 
    \item[d)] Marginal \textbf{box plots}, to identify the presence and amount of outliers for understanding the scales of the axes of the scatter plot. 
    \item[e)] Empirical copula \textbf{diagonals}, to visualize presence/absence of quadrant dependence, and possibly the need to partition the data (applying the gluing copula technique) to decompose the dependence into simpler quadrant dependencies.
    \item[f)] A bar chart showing the absolute value of the empirical Spearman's \textbf{concordance} $\rho_n$, and the Schweizer-Wolff's \textbf{dependence} measure $\sigma_n$ (if the bar colors are different then the sign of $\rho_n$ is negative). The combination of these two values is helpful for quantifying the degree of quadrant dependence, and for finding gluing points when applying the gluing copula technique.
\end{itemize}


In this paper we have organized these components into a $3\times 3$ grid, as shown in \cref{fig:dplot} and other examples (although other configurations would also be valid), which we call a \textit{dependence plot} (dplot). In the remainder of this section we will introduce two theoretical cases, while \cref{sec:realdata} will contain examples with real data.

\subsection{Example 1: Non quadrant dependence}
\label{sec:ex1}

\begin{figure}[t]
  \centering
  \includegraphics[width=\columnwidth]{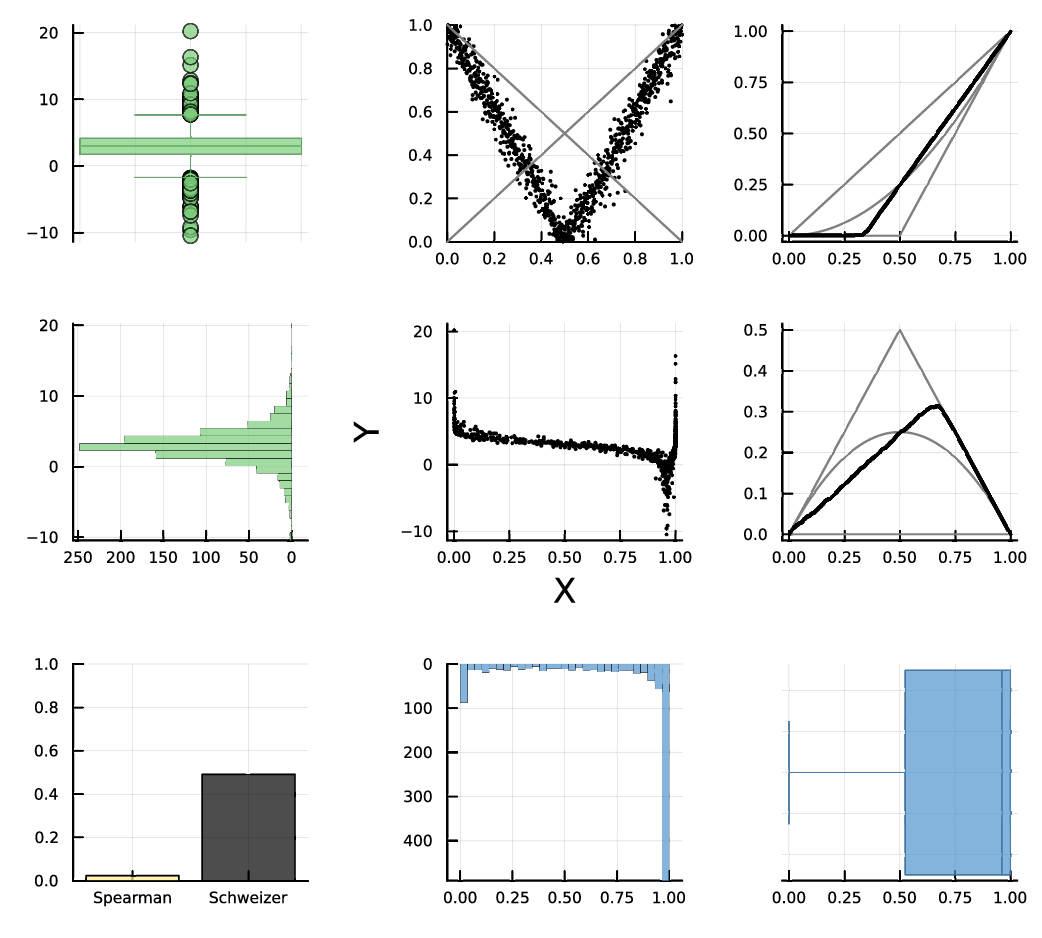}
  \caption{Dplot for Example 1 in \cref{sec:ex1}. \textit{Center:} Scatter plot of simulated values. \textit{Top-center:} Rank plot. \textit{Top-right and center-right:} Main and secondary diagonal sections of the empirical copula. \textit{Bottom-left:} Bar chart of the absolute value of the empirical Spearman's concordance ($|\rho_n|=0.02$) and the empirical Schweizer-Wolff's dependence ($\sigma_n=0.49$). If Spearman's bar is yellow then the sign of $\rho_n$ is negative, otherwise it is positive. \textit{Bottom-center and bottom right:} Histogram and box plot of $X$ values. \textit{Center-left and top-left:} Histogram and box plot of $Y$ values.}
  \label{fig:dplot}
\end{figure}

Consider a vector of two continuous random variables $(X,Y)$ with the following characteristics:
\begin{itemize}
    \item $X$ is bimodal (Kumaraswamy distribution with parameters $0.25$ and $0.15$);
    \item $Y$ is non-monotone unimodal (t-Student with location parameter $3.0$, scale parameter $1.5$, and $2.5$ degrees of freedom);
    \item If $X$ is below its median then there is NQD$(X,Y)$, while otherwise there is PQD$(X,Y)$. In both cases we have used the parametric Frank family of copulas (see \cite{Nelsen2006}) with parameters $-30.0$ and $30.0$, respectively.
\end{itemize}
This example falls into category R8 of \cref{tab:deptypes}, and its corresponding dplot is shown in \cref{fig:dplot}. In the center we have the usual scatter plot of $(X,Y)$. The associated rank plot is immediately above, and clearly exhibits a gluing of a NQD copula with a PQD one, with $\theta=0.5$ as gluing point. This can be confirmed by analyzing the graphs associated with the diagonal sections of the empirical copula with the independence copula $\Pi$ (top-right and center-right). Note that the empirical diagonals cross the $\Pi$ diagonals at $0.5$. Also, the empirical diagonals start being below $\Pi$ (which confirms NQD), but end up above $\Pi$ after $u=0.5$.

In the bar chart in the lower left corner we represent the absolute value of Spearman's empirical concordance together with Schweizer-Wolff's $\sigma_n$ (since it is useful to compare these magnitudes). Also, we indicate the sign of $\rho_n$ through color. For negative values we use a light color (in this case, yellow), and for positive values we use black (which we also use for the bar associated with $\sigma_n$). In this case, the bar chart shows a clear numerical difference $0.02=|\rho_n|<\sigma_n=0.49$ (by construction of the example, the theoretical value of Spearman's concordance is exactly equal to zero). This confirms that the dependence between $X$ and $Y$ is neither PQD nor NQD all the time, as expected.

The remainder of the graphs are related to marginal characteristics of the variables. There are no observed outliers for $X$, but several for $Y$. Also, the pdf for $X$ is considerably left-skewed. These are the main reasons why the dependency structure is hard to visualize in the scatter plot, but apparent in the marginal-free rank plot. Furthermore, users might erroneously perceive an overall decreasing trend since the points exhibit a slightly decreasing trend for $X\leq 0.9$. Moreover, the empirical Pearson's correlation coefficient in this case is $r_n=-0.40,$ which is quite misleading.


\subsection{Example 2: Positive quadrant dependence with noise}
\label{sec:ex2}

\begin{figure}[t]
  \centering
  \includegraphics[width=\columnwidth]{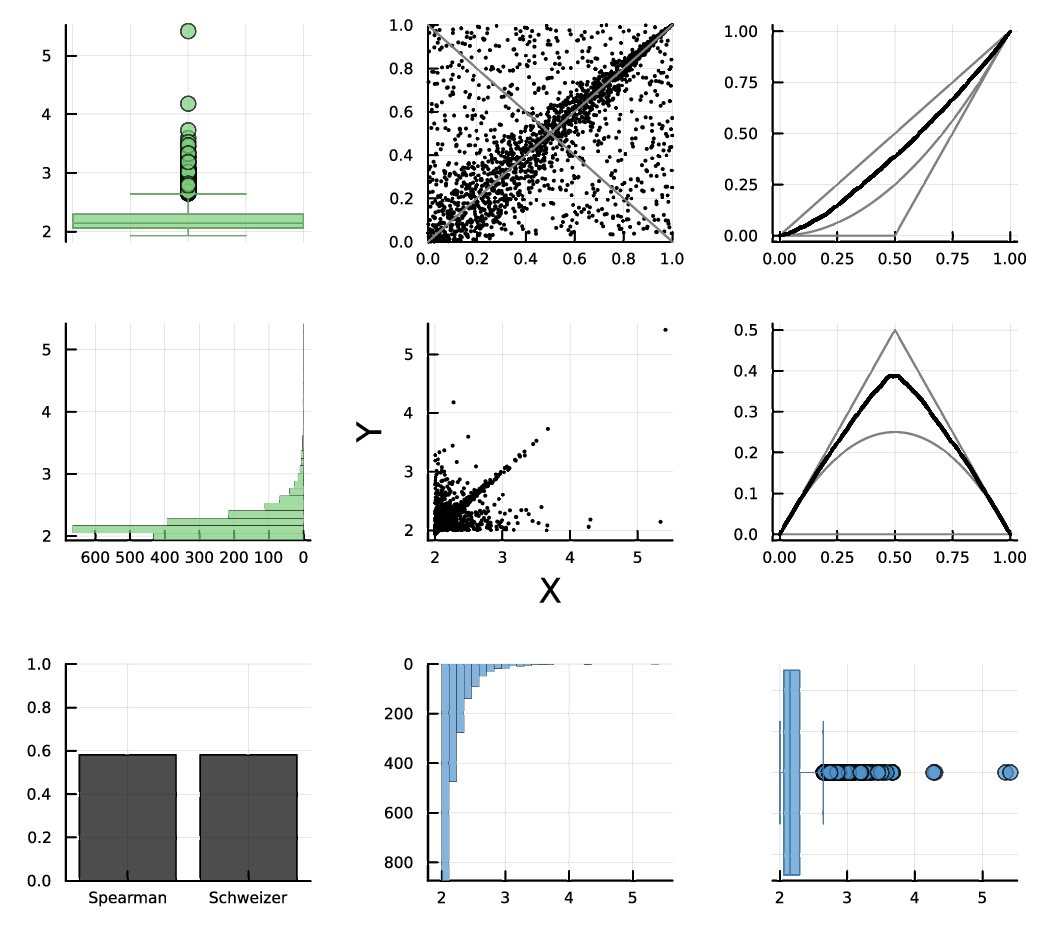}
  \caption{Dplot for Example 2 in \cref{sec:ex2}, where the rank plot illustrates the combination of PQD and independence. Thus, it falls into category R5 of \cref{tab:deptypes} and \cref{fig:rankplots}.}
  \label{fig:dplot2}
\end{figure}

Consider a random vector $(X,Y)$ where $X$ is a Pareto$(2,10)$ random variable, $\varepsilon$ a random noise distributed Normal$(0,0.03),$ $Z$ another Pareto$(2,10)$ random variable independent from $X,$ and $B$ a Bernoulli$(0.4)$ random variable. If we define random variable $Y$ as follows:
\[
    Y = (1-B)(X+\varepsilon) + BZ
\]
we have a probabilistic model for $(X,Y)$ where, with probability $0.6,$ they exhibit a strong linear relationship, but with probability $0.4$ they behave as independent random variables. \Cref{fig:dplot2} shows the corresponding dplot, where the rank plot is of type R5 (see \cref{tab:deptypes} and \cref{fig:rankplots}). Note that it can be interpreted as a (convex) combination of R1 (independence) and R2 (PQD, due to the positive linear relationship). In this example Schweizer-Wolff's dependence is $\sigma_n=0.58$, which is equal in magnitude and sign to the empirical Spearman's concordance, which implies PQD (see property B3 in \cref{subsec:depcon}). In this example, the diagonal sections also distinctly confirm PQD, as the empirical copula consistently exceeds the independence copula $\Pi$. Lastly, while discerning the linear relationship in the scatter plot is relatively straightforward, the independence relation is less apparent. Conversely, both relationships are clearly evident in the rank plot.

\section{Real data examples}
\label{sec:realdata}
In this section we illustrate the previous ideas and proposals with real data available from public sources.

\begin{figure}[t]
  \centering
  \includegraphics[width=\columnwidth]{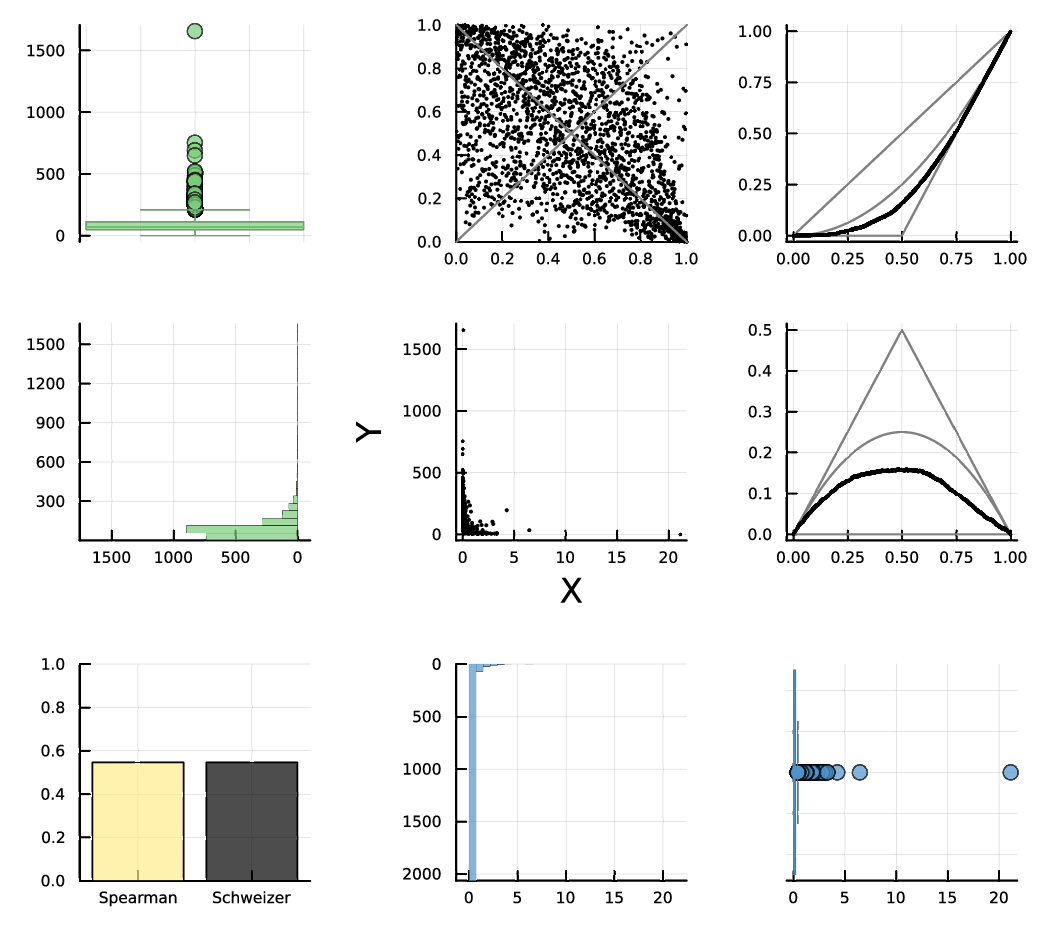}
  \caption{Dplot of the data set from \cite{Salim23} to analyze the dependence between the ``pay ratio'' ($X$) versus ``median worker pay'' ($Y$) in top 3,000 United States companies.}
  \label{fig:ex3}
\end{figure}

\subsection{Example 3: NQD with outliers}
\label{subsec:ex1}
In this example we use the data set ``CEO vs Worker Pay in Top 3000 US Companies [2023]'' (see \cite{Salim23}) to analyze the dependence between the ``pay ratio'' ($X=$ CEO/worker salary ) versus ``median worker pay'' ($Y$), see \cref{fig:ex3}. Firstly, it is difficult to analyze the relationship through the scatter plot, mainly due to the presence of outliers, which are clearly shown in the box plots. In this case, the rank plot falls into the R3 category (since the distribution of points is not uniform we can discard independence). Other components of the dplot also confirm a negative quadrant dependence. The equal height of the bars for Schweizer-Wolff's dependence and the absolute value of Spearman's concordance, where the latter is negative (yellow bar), along with the empirical diagonals that in both cases are below the independence diagonals, clearly indicate NQD. Numerically, $\sigma_n=0.55$ and $\rho_n=-0.55,$ in contrast with Pearson's correlation $r_n=-0.18,$ which is far from the concordance value since it is affected by the presence of outliers.


\subsection{Example 4: Gluing PQD and NQD}
\label{subsec:ex2}

\begin{figure}
  \centering
  \includegraphics[width=\columnwidth]{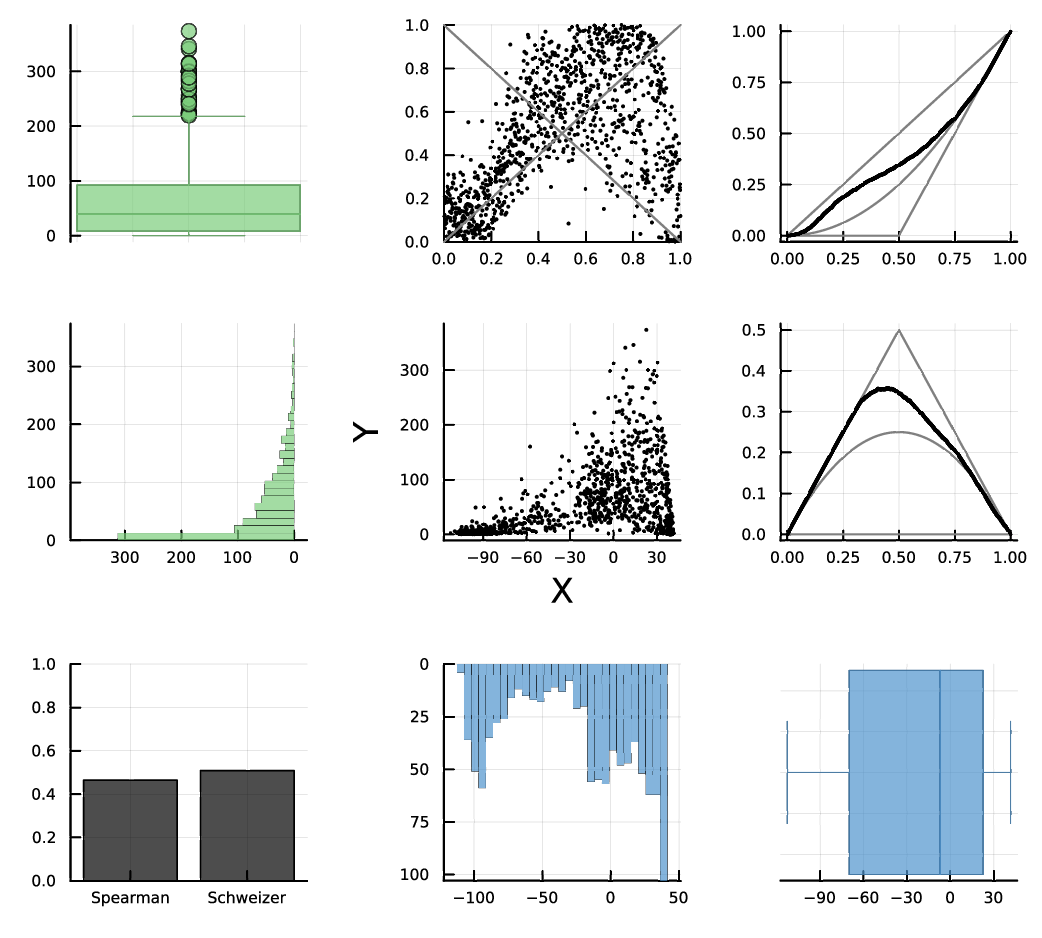}
  \caption{Dplot for variables ``infrared minimum value'' ($X$) and ``contrast'' ($Y$) of the ``Cloud'' data set~\cite{Collard89}.}
  \label{fig:ex4}
\end{figure}

For the next example we use the ``Cloud'' data set~\cite{Collard89} with features about images related to climate. Specifically, we analyze the dependence between ``infrared minimum value (ir-min)`` ($X$) and ``contrast'' ($Y$) through the dplot in \cref{fig:ex4}. The scatter plot appears to show a strictly increasing trend for $Y$ in terms of $X$ plus some random noise. However, the rank plot is suggesting something slightly different: an increasing trend for most part of the data, but decreasing for larger values of $X.$ This would explain the difference between Spearman's concordance $\rho_n=+0.46$ and Schweizer-Wolff's dependence $\sigma_n=0.51,$  which suggests the presence of both PQD and NQD. The empirical diagonals also indicate both PQD and NQD since they cross the independent copula $\Pi$ at about $u=0.8$, passing from being above $\Pi$ to lying slightly below it (in this case it may be necessary to zoom in).

\begin{figure}
  \centering
  \includegraphics[width=\columnwidth]{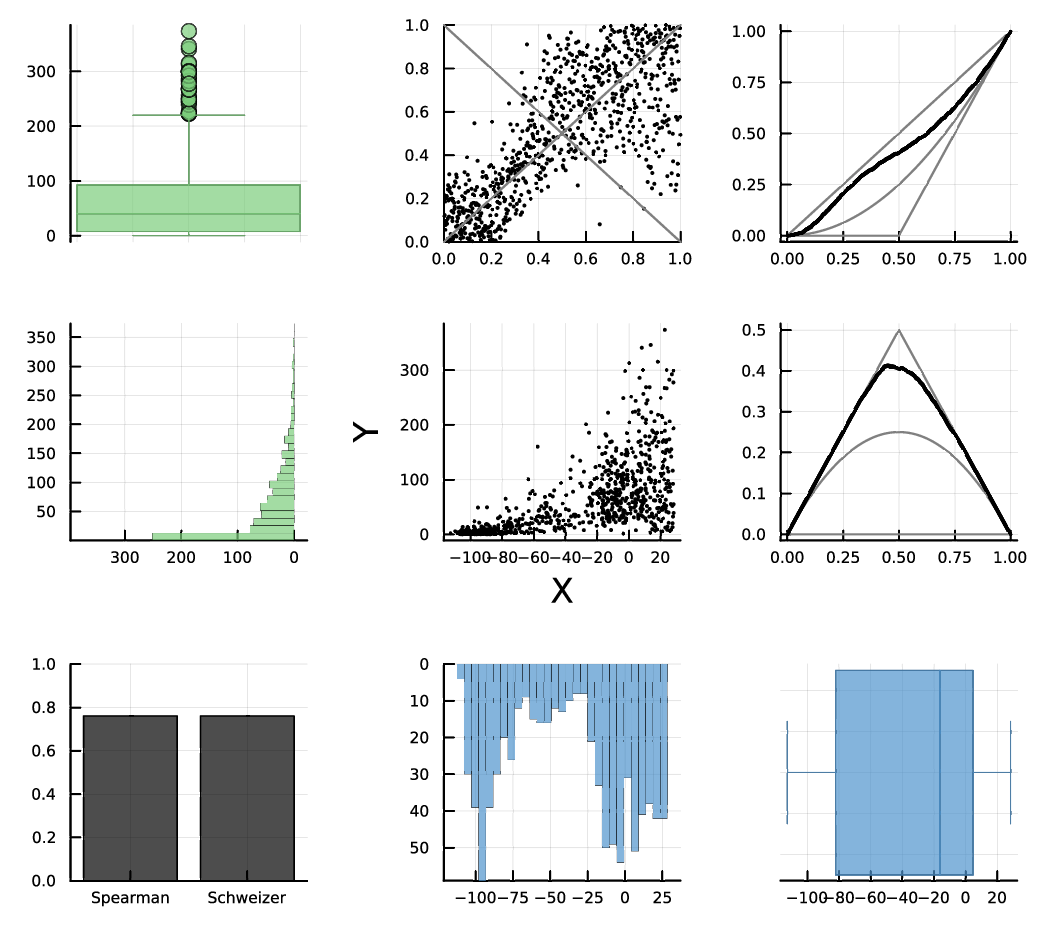}
  \caption{Dplot for variables ``infrared minimum value'' ($X$) and ``contrast'' ($Y$) of the ``Cloud'' data set~\cite{Collard89}, but conditioning on $X\leq F_n^{-1}(0.8)=28.36.$}
  \label{fig:ex4A}
\end{figure}

\begin{figure}
  \centering
  \includegraphics[width=\columnwidth]{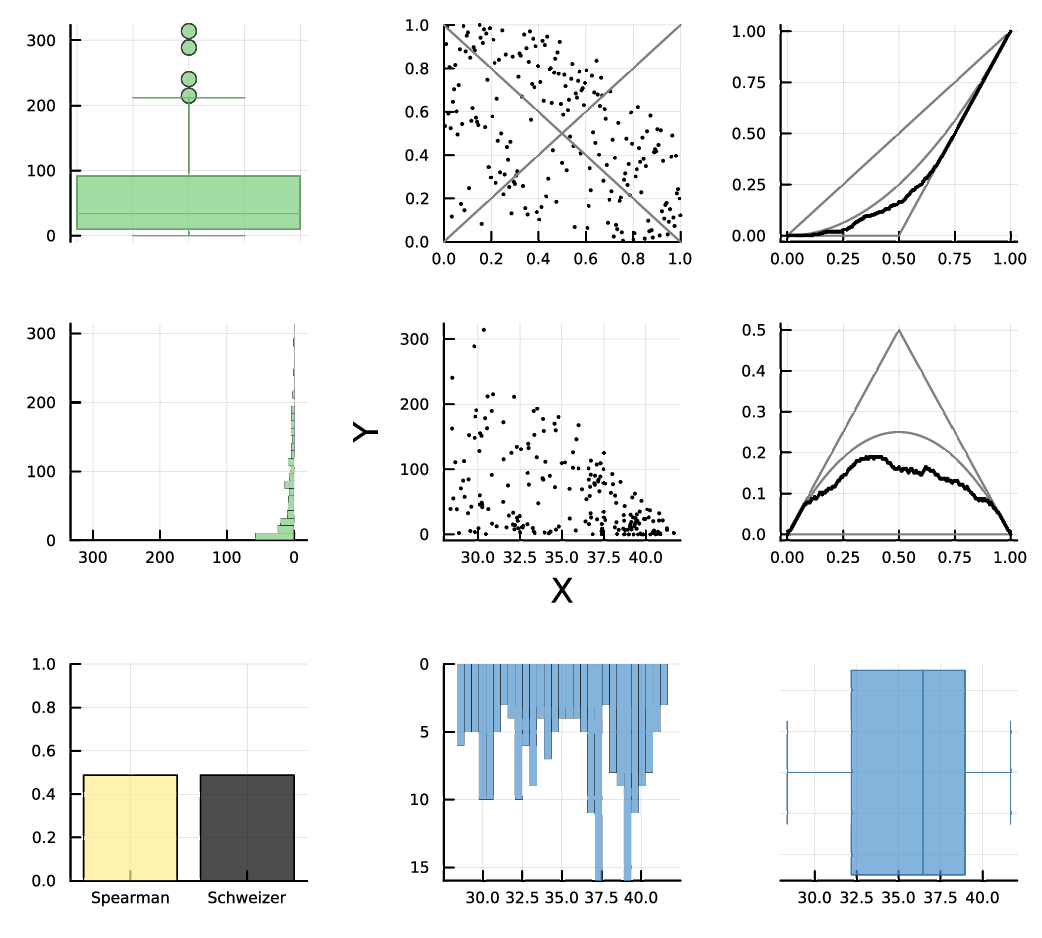}
  \caption{Dplot for variables ``infrared minimum value'' ($X$) and ``contrast'' ($Y$) of the ``Cloud'' data set~\cite{Collard89}, but conditioning on $X\geq F_n^{-1}(0.8)=28.36.$}
  \label{fig:ex4B}
\end{figure}

The analysis of the dplot suggests a gluing of a PQD copula followed by a NQD copula. If we order the bivariate observations in terms of the observed values of $X$ and split them into two subsets: the first 80\% as data subset 1, and the rest as subset 2, the corresponding dplots (see Fig.~\ref{fig:ex4A} and \ref{fig:ex4B}) reveal that subset 1 is PQD and subset 2 is NQD, which confirms that using $u=0.8$ as gluing point is a good choice. It is worth noticing that the best gluing point is the one such that in each subset the absolute value of Spearman's concordance equals that of Schweizer-Wolff's dependence. Specifically, for subset 1 we have $\rho_n^{(1)}=+0.76=\sigma_n^{(1)}$, and for subset 2 we have $\rho_n^{(2)}=-0.49$ and $\sigma_n^{(2)}=0.49.$

Finally, this would be an example of type R7 dependence, with gluing point $u=0.8$ of PQD followed by NQD. In terms of the original variables it is equivalent to splitting the observed data for $(X,Y)$ by conditioning $X\leq F_n^{-1}(0.8)=28.36$ for PQD, and $X\geq F_n^{-1}(0.8)=28.36$ for NQD. However, note that it is not easy to determine when the relationship between the variables begins to decrease in the scatter plot.

Jointly, or separately in subsets 1 and 2, the marginal density for $Y$ seems to be monotone unimodal, while $X$ appears to be multimodal, especially in Figures~\ref{fig:ex4} and \ref{fig:ex4A}, which produces the slight impression in the scatter plot of two clusters divided approximately by the vertical line $X=-30.$ Thus, the scenario is similar to the one in the graph in row 2, column 4, of \cref{fig:indeplots_new}. The appearance of clusters is due to the bimodal marginal distribution of $X$, and not to the statistical dependence between the variables.

\subsection{Example 5: Apparently independent}
\label{subsec:ex5}

\begin{figure}
  \centering
  \includegraphics[width=\columnwidth]{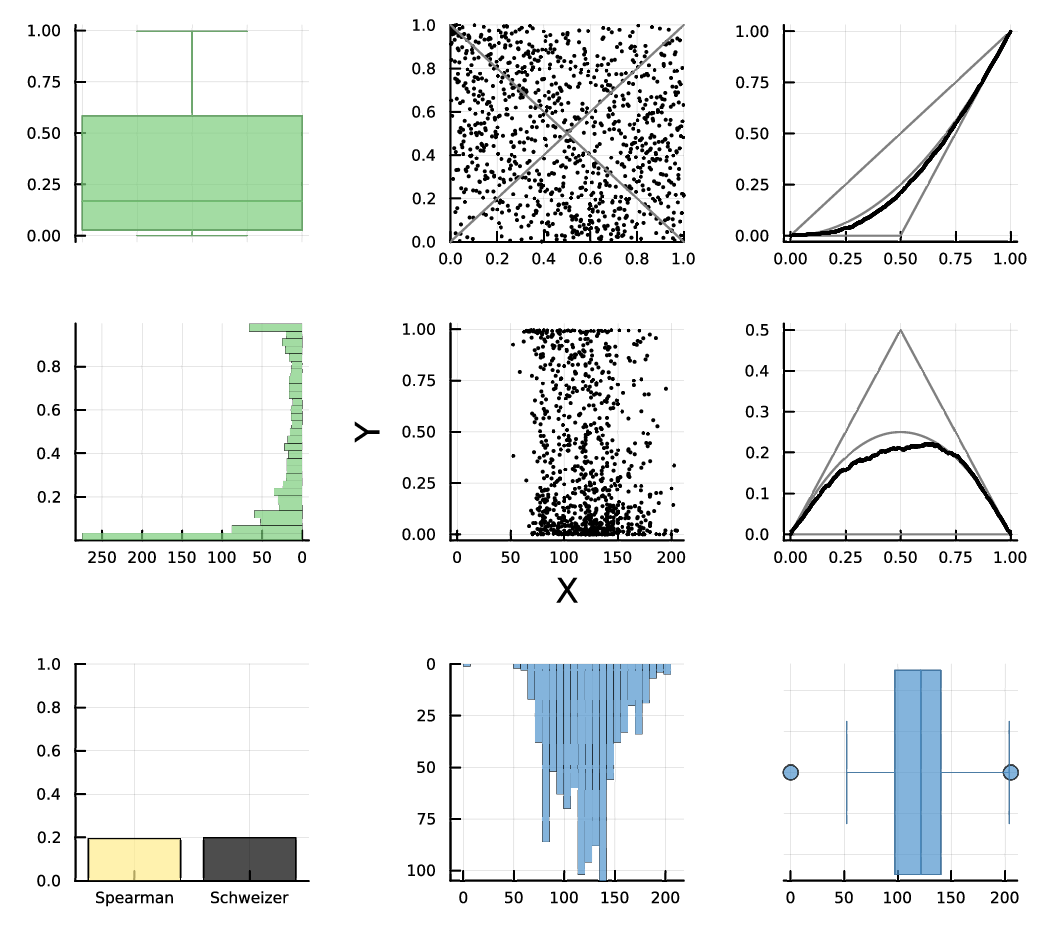}
  \caption{Dplot of a random subsample of size 1,000 from the ``Song Features Dataset - Regressing Popularity'' data set in \cite{Oturkar23}, comparing ``tempo'' ($X$) and ``acousticness'' ($Y$).}
  \label{fig:ex5}
\end{figure}

In this example we use a random subsample of size 1,000 from the ``Song Features Dataset - Regressing Popularity'' (Spotify Song features) data set in \cite{Oturkar23}, comparing tempo ($X$) and acousticness ($Y$), through the dplot in \cref{fig:ex5}. This is an example where from both the scatter plot, and even the rank plot, we may probably be tempted to assess independence at a first glance. For example, the scatter plot appears to be similar to the one in row 5, column 3, of \cref{fig:indeplots_new}. However, note a slight shift to the left of the points in the upper part of the scatter plot, compared to the points on the lower part. Also, in the rank plot there is slightly less density of points on the bottom-left and upper-right corners, compared to the density in the upper-left and bottom-right counterparts. Thus, the plots suggest a weak decreasing relationship. This is confirmed by examining the bar chart and diagonal sections. Specifically, not only does Schweizer-Wolff's dependence measure $\sigma_n=0.2$ indicate some dependence (albeit a weak one), in this case $\sigma_n = |\rho_n|$, with $\rho_n<0$, which clearly indicates NQD. Furthermore, the empirical diagonals also suggest NQD, since they lie (slightly) are below their independence counterparts. As a conclusion, this would be an example of weak dependence of type R3 according to \cref{tab:deptypes}, where by ``weak'' we mean not to far from R1, but not as clear as the R3 type in \cref{fig:ex3}. 

\section{Conclusions and discussion}
\label{sec:conclusions}

This work presents several contributions related to the visual assessment of dependence between two continuous random variables. Firstly, it reviews essential theory of copulas, necessary for understanding why rank plots should be chosen over traditional scatter plots for assessing dependence. In contrast to scatter plots, rank plots do not include uninformative, and possibly misleading, information about marginal distributions that are unrelated to dependence. The paper also provides guidelines (see \cref{tab:deptypes} and \cref{fig:rankplots}) for using and interpreting rank plots, identifying nine categories related to broad types of dependencies and combinations of these.

Regarding association measures, we have highlighted the superior reliability of dependence measures, such as Schweizer-Wolff's~(\ref{eq:schweizer}), and concordance measures like Spearman's~(\ref{eq:spearman}), over Pearson's correlation coefficient. Similarly to the comparison between rank plots and scatter plots, the key characteristic of Schweizer-Wolff's dependence and Spearman's concordance is that they derive solely from the copula, and are therefore unaffected by marginal distributions. Instead, Pearson's correlation combines information from the copula with marginal characteristics, and can therefore be misleading, as we have shown in several examples. We believe this is relevant for the entire scientific community, since Pearson's correlation is arguably the most used association measure used in practice, and the default option in many software packages, despite previous efforts to communicate its limitations (see \cite{Embrechts1999}). It is also relevant for the visualization community since Pearson's correlation has been studied intensively in relation to scatter plots. Moreover, some authors may view scatter plots as tools for communicating Pearson correlation~\cite{Strain23a,Strain23b}.




The paper coalesces around the idea of the \textit{dependence plot} (dplot), an ensemble of nine graphs that encapsulates both the scatter plot and marginal distributions, together with visualizations that focus solely on aspects of dependence. The former are useful for detecting clusters, outliers, and examining the specific data values, among other tasks. Regarding the latter, rank plots provide a faithful description of the dependence between the variables, while Spearman's concordance and Schweizer-Wolff's dependence are appropriate association summaries, since they depend exclusively on the copula. In addition, the visualizations related to diagonal sections of the copula can help users decompose complex dependence patterns into simpler quadrant-dependent scenarios by conditioning on one variable. This idea is related to gluing copulas (i.e., describing the dependence through several copulas). Although we have addressed the gluing of two copulas, extending this concept to multiple copulas and/or conditioning in both variables is straightforward.





The ideas put forward in the paper can be useful for using or developing other visualization techniques in which it may be appropriate to visualize the dependency between two continuous random variables. For example, the theory clearly advocates for replacing data values with ranks (note that it would be straightforward to incorporate ranks in methods like parallel coordinates~\cite{Inselberg90}, table lens~\cite{Rao94}, and many others). Naturally, researchers should also consider replacing or complementing Pearson's correlation with Spearman's rank correlation or Schweizer-Wolff's dependence in their visualizations.


The general approach of this paper is non-parametric, as no model assumptions are made regarding the bivariate data under analysis. Nevertheless, the guidelines provided in \cref{tab:deptypes} for interpreting rank plots are useful in making decisions about fitting specific parametric families of copulas. For instance, if it is evident from the data that negative quadrant dependence (NQD) is present, then only parametric families of copulas accommodating NQD should be considered for goodness-of-fit testing. Similarly, if the rank plot of the data resembles, for example, case R8, the data should first be divided using the gluing copula technique. Subsequently, we may try fitting a parametric family of copulas with NQD to one subset, and a parametric family with positive quadrant dependence (PQD) to the other subset.

Moving forward, in future work we plan on exploring scenarios involving non-continuous random variables. While Sklar's Theorem remains applicable, the relationship between the joint distribution and its marginals transitions from a copula to a \textit{subcopula}, restricted to a proper subset of the unit square. In such contexts, traditional measures like those of Schweizer-Wolff and Spearman lose their validity, prompting the need for alternative measures, as discussed in~\cite{Erdely2017b}. Finally, we also envision carrying out perceptual studies of Spearman's rank correlation and Schweizer-Wolff's dependence measure on rank plots, given their superiority with respect to Pearson's correlation and scatter plots for assessing dependence.

\section{Reproducibility}
\label{sec:reproduce}

All the Julia programming code and data sets used for calculations and generating figures is available for reproducibility at \url{https://github.com/aerdely/visualdep}

\section*{Acknowledgements}

This work was partially supported by a grant from Universidad Nacional Aut\'onoma de M\'exico (PAPIIT IN104425) and by a grant from Universidad Rey Juan Carlos (2023/SOLCON-132212).


\bibliographystyle{abbrv-doi-hyperref}

\bibliography{VIS_2024}

\end{document}